\documentclass[%
 reprint,
superscriptaddress,
 amsmath,amssymb, 
 aip,
onecolumn,
]{revtex4-2}

\usepackage{graphicx}
\usepackage{dcolumn}
\usepackage{bm}
\usepackage{hyperref}
\usepackage{xcolor}
\usepackage{subfigure} 
\usepackage{setspace}
\usepackage{ulem}
\newcommand{\etal}{{\it et al.}}

\begin{document}

\title{Magnetohydrodynamic shock refraction at an inclined density interface}

\author{Fang Chen}
\affiliation{%
Mechanical Engineering, Physical Science and Engineering Division, King Abdullah University of Science and Technology, Thuwal 23955-6900, Saudi Arabia
}%
\author{Vincent Wheatley}
\affiliation{
Mechanical and Mining Engineering, University of Queensland, St Lucia QLD 4072, Australia
}%
\author{Ravi Samtaney}%
 \email{ravi.samtaney@kaust.edu.sa} 
\affiliation{%
Mechanical Engineering, Physical Science and Engineering Division, King Abdullah University of Science and Technology, Thuwal 23955-6900, Saudi Arabia
}%
\date{\today}
\begin{abstract}
Shock wave refraction at a sharp density interface is a classical problem in hydrodynamics. 
Presently, we investigate the strongly planar refraction of a magnetohydrodynamic (MHD) shock wave at an inclined density interface. A magnetic field is applied that is initially oriented either perpendicular and parallel to the motion of incident shock.  
We explore flow structure by varying the magnitude of the magnetic field governed by the non-dimensional parameter $\beta \in (0.5, 10^6)$ and the inclination angle of density interface $\alpha \in (0.30, 1.52)$. 
The regular MHD shock refraction process results in a pair of outer fast shocks (reflected and transmitted) and a set of inner nonlinear magneto-sonic waves. 
By varying magnetic field (strength and direction) and inclination interface angle, the latter waves can be slow shocks, slow expansion fans, intermediate shocks or slow-mode compound waves. 
For a chosen incident shock strength, and density ratio, the MHD shock refraction transitions from regular (all nonlinear waves meeting at a single point) into irregular when the inclined density interface angle is less than a critical value. 
Since the MHD shock refraction is self-similar, we further explore by converting the initial value problem (IVP) into a boundary value problem (BVP) by a self-similar coordinate transformation. The self-similar solution to the BVP is numerically solved using an iterative method, and implemented using the p4est adaptive mesh framework. 
The simulation shows that a Mach stem occurs in irregular MHD shock refraction, the flow structure can be an MHD equivalent to a single Mach reflexion irregular refraction $MRR$ and convex-forwards irregular refraction $CFR$ that occur in hydrodynamic case. 
For Mach number $M = 2$, both analytical and numerical results show that perpendicular magnetic fields suppress the regular to irregular transition compared to the corresponding hydrodynamic case.
As Mach number decreased, it is possible that strong perpendicular magnetics promote the regular to irregular transition while moderate perpendicular magnetics suppress this transition compared to the corresponding hydrodynamic case.
\end{abstract}

\maketitle
\normalem
\section{\label{sec:1} Introduction}
\begin{figure}[] 
\includegraphics[scale=0.10]{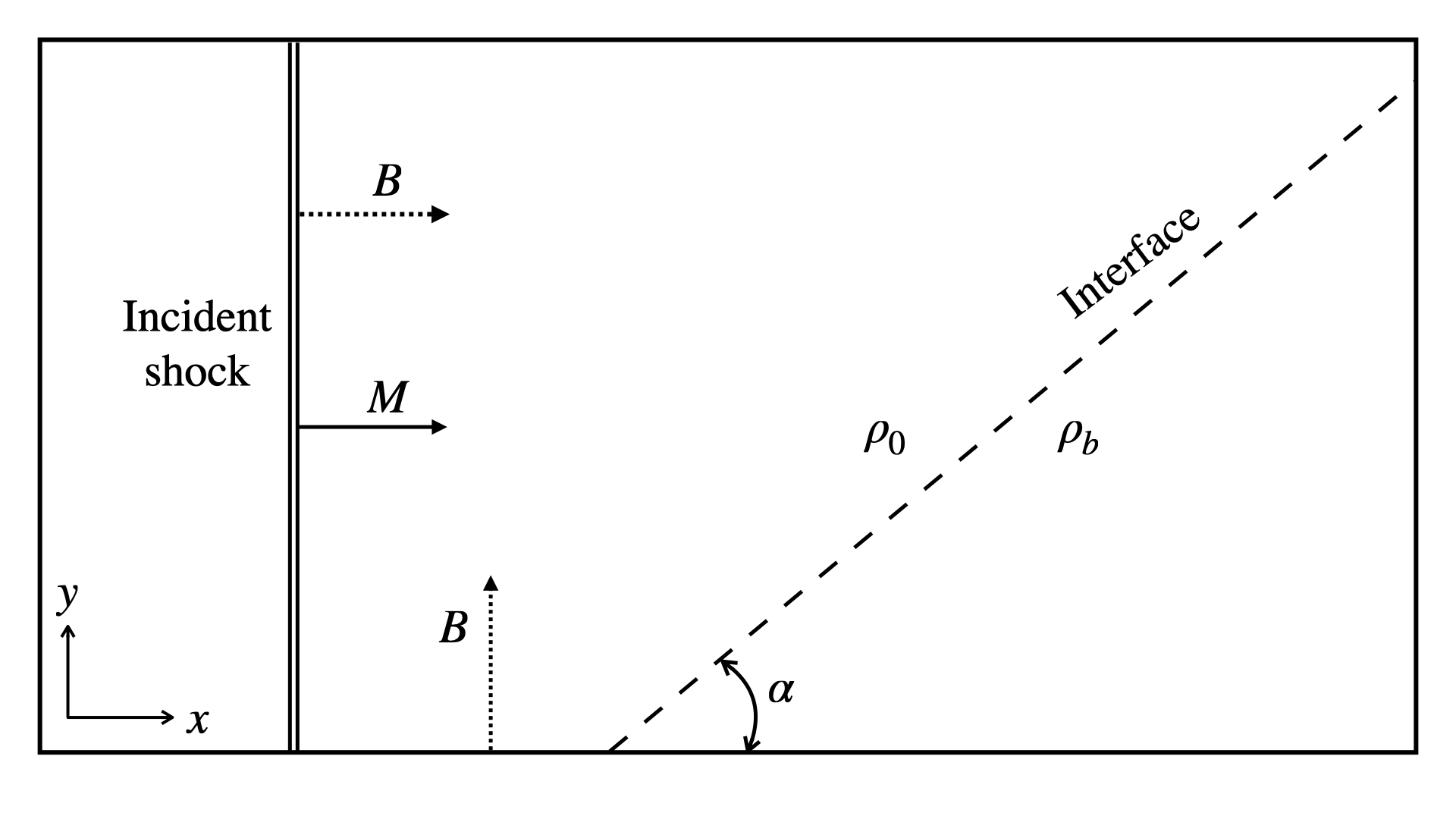}
\caption{\label{fig:1.1} Physical set-up for the problem of shock refraction. The initial pressure  in the unshocked regions is $p_0 =1$. The magnetic field could be absent, parallel or perpendicular to the motion of incident shock.} 
\end{figure}

Understanding magnetohydrodynamic (MHD) shock refraction is important for any application involving shock waves and variable density flows, such as inertial confinement fusion (ICF)~\cite{Lindl2004}, as well as astrophysical phenomena~\cite{Arnett2000}, etc. 
A canonical physical set-up to investigate shock refraction  is shown in Fig.~\ref{fig:1.1}. 
The flow is characterized by the parameters: the incident shock sonic Mach number $M$ (fast magnetosonic Mach number for fast mode MHD shocks), the density ratio of the interface $\eta = \rho_b / \rho_0$, the ratio of specific heats $\gamma$, the angle between the incident shock normal and interface $\alpha$, and the non-dimensional strength of the applied magnetic field $\beta^{-1} = B^2/{2p_0}$, where $B$ and $p_0$ denotes the dimensionless magnitude of the applied magnetic field and the pressure of the gas.  

In hydrodynamic, the regular shock refraction results in a transmitted $T$ and a reflected $R$ shock~\cite{Henderson1978, henderson1989}, on the other hand, a Mach stem is observed and the shock refraction becomes irregular as the inclination angle of interface is decreased less than critical angle~\cite{henderson1968,Henderson1991}. 
Where all waves resulting from the refraction process meet at a point (triple point) and are planar, this is known as regular shock refraction, otherwise, the shock refraction is irregular.
Experiments have shown that the irregular refraction structure includes centered expansion type of refraction $CER$, single Mach reflexion type of refraction $MRR$ and concave-forwards type of refraction $CFR$~\cite{Henderson1978}, etc.
 
\begin{figure}[]   
\centering
\subfigure[Regular MHD refraction]{
\includegraphics[scale=0.15]{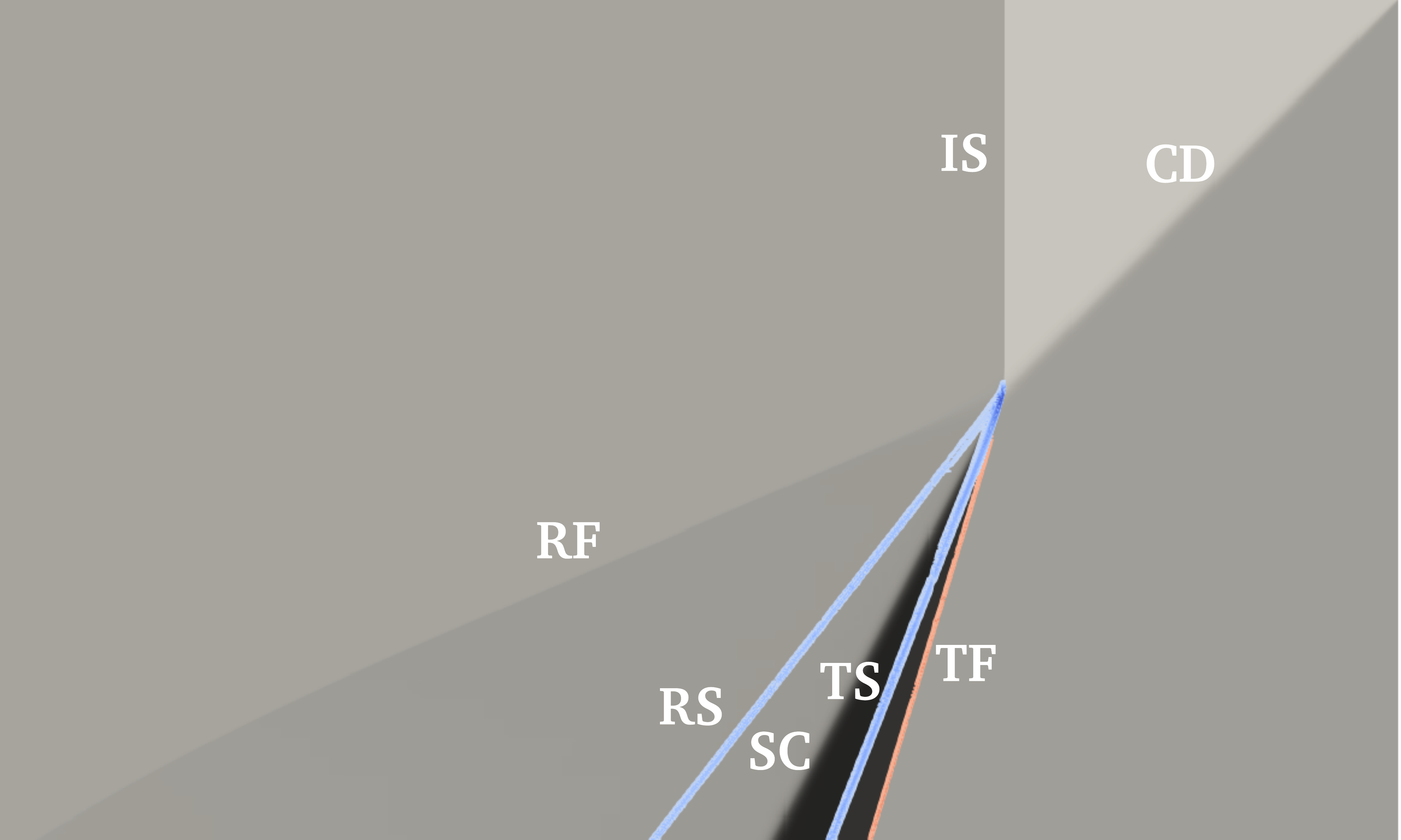}\label{fig:1.2a}
} 
\quad
\subfigure[Irregular MHD refraction]{ 
\includegraphics[scale=0.15]{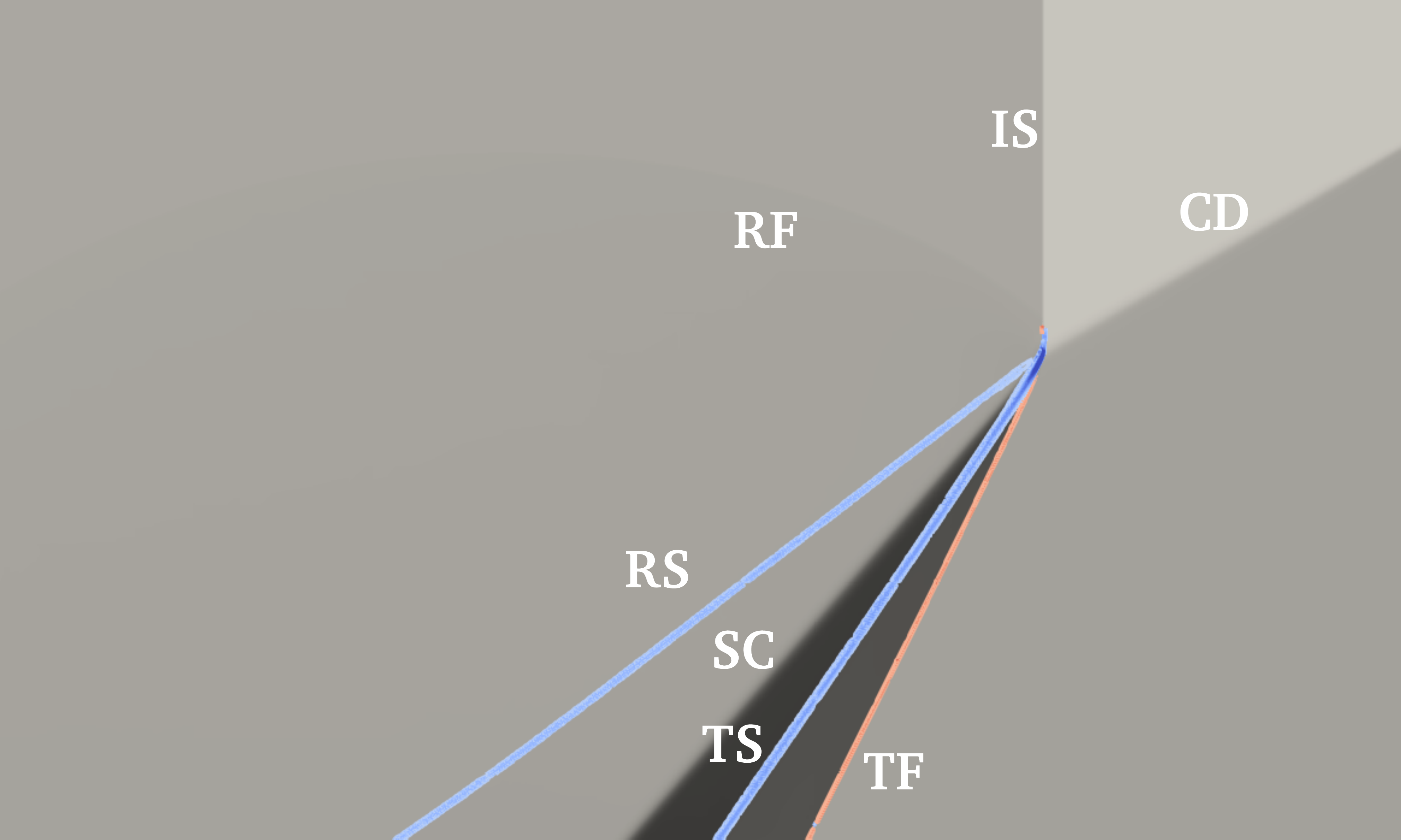}\label{fig:1.2b}
} 
\quad
\caption{Numerical density field overlaid with vorticity resulting from a MHD shock refraction in presence of a magnetic field which is parallel to the motion of incident shock, with $M = 2, \eta = 3, \gamma = 1.4, \beta = 2 $, (a) $ \alpha = \pi/4$, (b) $ \alpha = \pi/6$. $IS$, incident shock; $CD$, contact discontinuity; $TF$, transmitted fast shock; $TS$, transmitted slow/intermediate wave; $SC$, shocked contact; $RS$, reflected slow/intermediate wave; $RF$, reflected fast wave.}
\label{fig:1.2}       
\end{figure}

For strongly planar ideal MHD, Samtaney~\cite{Samtaney2003} numerically investigated the shock refraction with the presence of a magnetic field which is initially parallel to the motion of incident shock. 
The problem is characterized with $M = 2, \eta = 3, \gamma = 1.4, \alpha = \pi/4$ and $\beta = 2$, see in Fig.~\ref{fig:1.1}. 
Here, we define a flow to be {\em planar} if there is no derivatives in the out-of-plane $(z)$ direction, and {\em strongly planar} if there is also a reference frame in which there is no vector component in the $z$-direction. 
In this study, there are a pair of reflected shocks and a pair of transmitted shocks as shown in Fig.~\ref{fig:1.2a}. $RS$ and $RF$ are the slow reflected and fast reflected magneto-sonic shocks, respectively; whereas $TS$ (slow or intermediate shock) and $TF$ (fast shock) are the transmitted magneto-sonic shocks. 
Samtaney noted  that the growth of the Richtmyer-Meshkov instability is suppressed in the presence of a magnetic field, since the vorticity is transported away from the density interface onto a pair of slow or intermediate magneto-sonic shocks $(RS$ and $TS)$. 
Consequently, the density interface is devoid of vorticity and its growth and associated mixing is suppressed. 
Based on this case~\cite{Samtaney2003}, Wheatley \etal ~\cite{Wheatley2005JFM} developed analytical solutions to the planar and strongly planar regular MHD shock refraction by fixing inclination angle $\alpha = \pi/4$ and varying $\beta \in (2, 10^7)$. They focused on {\em regular} refraction where all the waves meet at a single point. Furthermore,
they showed that the resulting refraction structure might consist of five, six or seven waves, which depend on the different wave types of the kind. The wave pattern may include fast, intermediate, and slow MHD shocks, slow compound waves, $180^{\circ}$ rotational discontinuities $(RDs)$, and slow-mode expansion fans. 
However, there is no analytical solution for irregular MHD shock refraction at present. 
In Fig.~\ref{fig:1.2b}, we show the case of an irregular MHD shock refraction obtained numerically with the same parameters as the case in~\cite{Samtaney2003} except $\alpha = \pi/6$. A Mach stem separates $RF$ and $IS$ from the point that the other waves meet and the resulting refraction wave pattern is irregular. 

One of the main challenges in the MHD shock refraction is the appearance of inadmissible waves, especially those of the intermediate type. 
In the three-dimensional MHD system of equations,  the evolutionary condition~\cite{Akhiezer1959, Jeffrey1964, Polovin1990} restricts physically admissible discontinuities to fast shocks, slow shocks, contact discontinuities and $180^{\circ} RDs$ . 
The evolutionary of intermediate shocks has been extensively studied and is somewhat controversial. 
Falle and Komissarov~\cite{falle2001} demonstrated that a shock is physical only if it satisfies both the viscosity admissibility condition and the evolutionary condition. 
In this framework, for a planar system fast and slow shocks are evolutionary and have unique structurally stable dissipative structures, as well as $RDs$ are considered admissible, while all intermediate shocks are non-evolutionary and can be destroyed by interactions with Alfv\'en waves. 
This is in contradiction to the conclusion obtained via numerical tests by Wu~\cite{wu1990,wu1994,wu1995}. 
For a strongly planar system, Falle and Komissarov~\cite{falle2001} concluded that $1\to 3$ and $2\to 4$ intermediate shocks along with slow (noted as $C_1$) and fast (noted as $C_2$) compound waves are shown to be evolutionary and have unique dissipative structures. 
Both $2\to 3$ intermediate shocks and $180^{\circ} RDs$ are found to be non-evolutionary. 
These results are in agreement with those of Myong and Roe~\cite{myongroe1997}. We choose the conditions from the work of Falle and Komissarov due to its completeness in order to develop our MHD shock refraction analytical solutions. 

The previous work by Samtaney ~\cite{Samtaney2003} and Wheatley \etal ~\cite{Wheatley2005JFM} focused on regular refraction wherein the magnetic field is initially parallel to the direction of propagation of the incident shock. 
In present work, based on the method presented by Wheatley \etal ~\cite{Wheatley2005JFM}, we present analytical solutions to the problem of strongly planar regular shock refraction in the presence of a magnetic field which is initially perpendicular $\bm{B} = (0, B_y, 0)$ to the motion of incident shock $\bm{V} = (v_x, 0, 0)$ (hereinafter referred to as perpendicular magnetic field, whereas the case in~\cite{Samtaney2003} is referred as parallel magnetic field). 
Noting that the MHD shock refraction is self-similar, we further explore the phenomenon of MHD shock refraction by numerical simulations. For this, we first convert the equations of ideal MHD governing the initial value problem (IVP) into a boundary value problem (BVP) by a self-similar coordinate transformation. 
Main advantage is that  the IVP for inviscid flow problems with vortex sheets is ill-posed, whereas the self-similar BVP system is well-posed \cite{samtaney1996}. 
The IVP does not appear to converge to a weak solution as the mesh is refined, while the self-similar solution seems to converge with decreasing mesh size to a weak solution. 
The numerical results exhibit a very high resolution around discontinuities (here high-resolution refers to a sharper or higher gradient approximation to a discontinuity). An interesting outcome of the self-similar transformation concerns solutions involving contact slip lines (aka vortex sheets).
The self-similar transformation as well as the numerical method to solve the resulting BVP have been described in detail in the recent paper by Chen \& Samtaney~\cite{Chen2021}.  In this paper, the authors employed the Generalized Lagrangian multiplier GLM-MHD equations. 
Presently,  the self-similar GLM-MHD equations are employed to investigate both regular and irregular refractions. Moreover, two orientations, parallel and perpendicular to the incident shock propagation direction, are investigated numerically.
 
The outline of this paper is as follows. In Section~\ref{sec:2} we briefly introduce the numerical and analytical approaches. 
We present the results and discussion for regular refraction in Section~\ref{sec:3}, including different wave configurations resulting from varying strength of magnetic field $\beta$ and inclination angle of interface $\alpha$ focusing more on the solutions obtained by analytical means. In  Section~\ref{sec:4}, we present numerical simulations for irregular MHD shock refraction patterns and identify a couple of MHD shock refraction patterns that are similar to the MRR and CFR configurations noted in irregular shock refractions in hydrodynamics. 
 Finally, a summary is presented in the concluding Section~\ref{sec:5}.

\section{\label{sec:2} Equations, Analytical Method and Self-similar Formulation}
\subsection{Ideal MHD equations}
The ideal MHD equations, appropriately non-dimensionalized, in multiple spatial dimensions are 
\begin{subequations} 
\label{eq:2.00}
 \begin{align}
&\frac{\partial \rho}{\partial t} + \nabla \ . \  (\rho \boldsymbol{v}) = 0, \label{eq:1a}\\ 
&\frac{\partial (\rho\boldsymbol{v})}{\partial t} + \nabla \ . \ \Big[\rho\boldsymbol{vv}^{T} + \Big(p+\frac{\boldsymbol{B}.\boldsymbol{B}}{2}\Big)\boldsymbol{I} - \boldsymbol{BB}^{T}\Big] = \boldsymbol 0, \label{eq:1b}\\
&\frac{\partial \boldsymbol{B}}{\partial t} + \nabla \times (-\boldsymbol{v}\times \boldsymbol{B})  = \boldsymbol 0, \label{eq:1d}\\
&\frac{\partial E}{\partial t} + \nabla \ . \ \Big[\Big(E+p+\frac{\boldsymbol{B}.\boldsymbol{B}}{2}\Big)\boldsymbol{v} - (\boldsymbol{v} \ . \ \boldsymbol{B})\boldsymbol{B}\Big] =0,\label{eq:1c} \\
&\nabla. \  \boldsymbol{B} = 0. \label{eq:1e} 
\end{align}
\end{subequations}
Here $\rho, p $ and $ E $ are the fluid density, hydrodynamic pressure and total energy per unit volume, respectively;  $\boldsymbol{v} $ and $\boldsymbol{B} $ denotes the velocity vector and the magnetic induction. The ideal gas is considered in the present work, and the gas hydrodynamic pressure $p$ and the total pressure $p_t$ are given by 
 \begin{equation}
 \label{eq:2.01}
p=(\gamma - 1)\Big(E - \rho\frac{\boldsymbol{v}.\boldsymbol{v}}{2} - \frac{\boldsymbol{B}.\boldsymbol{B}}{2} \Big), \ \  \ \  \ \ p_t = p + \frac{\boldsymbol{B}.\boldsymbol{B}}{2},
\end{equation}
respectively. We will seek solutions to the strongly planar ideal MHD equations for two-dimensional (2D) conditions. 
\subsection{\label{subsec:2.1}Analytical method}
The MHD Rankine–Hugoniot (RH) relations govern weak solutions to the steady-state form of equations of ideal MHD (Eq.~\ref{eq:2.00}) corresponding to discontinuous changes from one state to another. We assume that all dependent variables vary only in the direction normal to the shock front, which is denoted with the subscript $n$. We also assume that all velocities and magnetic fields are coplanar, as we are seeking strongly planar ideal solutions. Under these assumptions, the set of jump relations for a stationary discontinuity separating two uniform states are~\cite{Sutton1965}:
\begin{subequations} 
\label{eq:2.11}
 \begin{align}
&[\rho v_n]= 0, \label{eq:2.11a}\\ 
&\Big[ \rho v_n^2 + p + \frac{B_t^2}{2} \Big]= 0, \label{eq:2.11b} \\ 
&[ \rho v_n v_t - B_n B_t ]= 0, \label{eq:2.11c}  \\ 
&\Big[ \frac{\rho v_n}{2} (v_n^2 + v_t^2) + \frac{\gamma v_n p}{\gamma - 1} + v_n B_t^2 - v_t B_n B_t   \Big] = 0,  \label{eq:2.11d}  \\ 
&[v_n B_t - v_t B_n] = 0.  \label{eq:2.11e}
\end{align}
\end{subequations}
Here, the subscript $t$ denotes the component of a vector tangential to the shock, and $[\mathcal X] = \mathcal X_2 -\mathcal X_1$ denotes the difference in the quantity $\mathcal X$ between the states upstream (subscript 1) and downstream (subscript 2) of the shock. We introduce a set of normalized variables~\cite{kennel1989} as follow: 
\[ r = \frac{v_{n2}}{v_{n1}}, \quad b = \frac{B_{t2}}{B_1},  \quad V_t = \frac{v_{t2}}{v_{n1}},  \quad \sin \theta_1 = \frac{B_{t1}}{B_{1}},  \]
where $\theta_1$ is the angle between the upstream magnetic field and the shock normal. We rewrite the Eq.~(\ref{eq:2.11}) to obtain the following algebraic equations in $r$ and $b$~\cite{Liberman1986}: 
\begin{subequations} 
\label{eq:2.12}
 \begin{align}
&\mathcal F(r,b) = Ar^2 + B(b)r + C(b) = 0, \label{eq:2.12a}\\ 
&Z(r,b) = bX - Y\sin\theta_1 = 0,  \label{eq:2.12b}
\end{align}
\end{subequations}
where detailed definitions of $A, B, C, X$ and $Y$ can be seen in Wheatley \etal ~\cite{Wheatley2005JFM}. The intersections of the curves defined by $\mathcal F = 0$ and $Z = 0$ are the locations in $(r, b)$ space where all jump conditions are satisfied. Subsequently, we can exactly calculate the downstream state of a discontinuity with given upstream state. 
\begin{figure}[b] 
\includegraphics[scale=0.8]{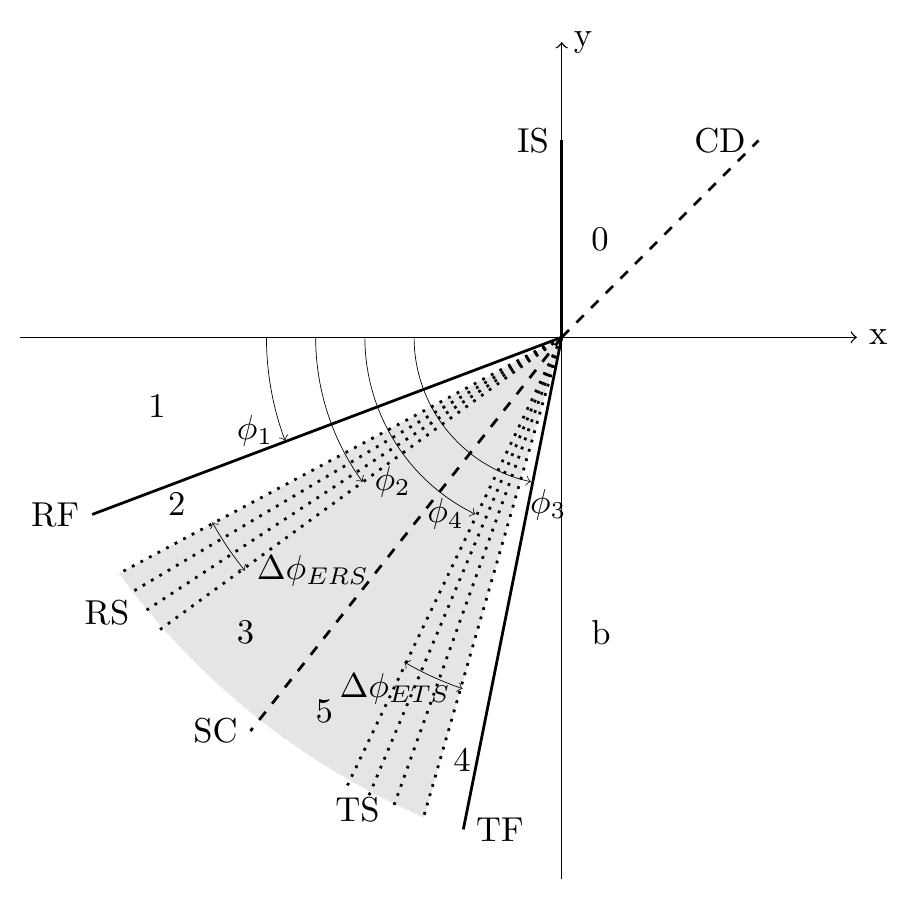}
\caption{\label{fig:2.1} Designations of the angles and regions of uniform flow for a shock refraction problem where $RS$ and $TS$ are slow-mode expansion fans. The undisturbed conditions to the left and right of the $CD$ are denoted states $0$ and $b$, respectively.
} 
\end{figure}
The wave configuration of the MHD shock refraction can be shown as in Fig.~\ref{fig:2.1}. 
There are four unknown angles $\phi_1, \phi_2, \phi_3$ and $\phi_4$ which identifies the location of $RF, RS, TF$ and $TS$, respectively. 
If discontinuity is an expansion fan or compound wave, $\phi$ defines the location of the trailing edge (or tail) of an expansion fan. 
The states 3 and 5 are the conditions to the left and right of the $SC$, then the following matching conditions must hold across the $SC$: 
\begin{subequations} 
\label{eq:2.13}
 \begin{align}
&p_3 = p_5, \label{eq:2.13a}\\ 
&v_{3x} = v_{5x},  \label{eq:2.13b}\\
&v_{3y} = v_{5y},  \label{eq:2.13c}\\
&|\bm{K}_3| = |\bm{K}_5|, \label{eq:2.13d}\\
&\bm{K}_3/|\bm{K}_3| =\bm{K}_5/ |\bm{K}_5|. \label{eq:2.13e}
\end{align}
\end{subequations}
Here, adopting the nomenclature from Wheatley \etal ~\cite{Wheatley2005JFM}, $|\bm{K} | = K = \beta^{-1/2}$, $K_n = K\cos\theta$, and $K_t = K\sin\theta$. The main procedure to find the solution is as follows.
First, we postulate a wave configuration including four plane waves. Mathematically this corresponds to selecting which root of the Rankin-Hugoniot relations will be used to compute the jumps across each shock.
Therefore, the guessed wave angles and the given state 0 and $b$ allow us to compute the conditions on either side of the SC. An approximate solution to the strongly planar MHD shock refraction problem is then obtained by iterating on the wave angles until the matching conditions Eq.~(\ref{eq:2.13}) are satisfied to six significant figures. The full solution technique for the MHD regular shock refraction problem is documented in~\cite{Wheatley2005JFM}.
In the strongly planar system, note that the fast and slow waves (shocks or expansion fans), contact discontinuities, $1 \to 3$ and $2 \to 4$ intermediate shocks, slow and fast compound waves are considered admissible according to   Falle and Komissarov~\cite{falle2001}. 

\subsection{\label{subsec:2.2}Self-similar GLM-MHD method}
In the approach of Dedner \etal ~\cite{Dedner2002}, the divergence constraint of the magnetic field (Gauss’s law) is coupled to Faraday’s equation by introducing a new scalar field function or generalized Lagrangian multiplier (GLM) $\psi$. The resulting GLM-MHD system in full dimensional system $\boldsymbol{x} = (x, y, z)$ can be rewritten in conservative form as 
 \begin{equation}
 \label{eq:2.21}
 \frac{\partial \boldsymbol{U}}{\partial t} + \sum_{j=x,y,z}  \frac{\partial \boldsymbol{F}_j (\boldsymbol{U})}{\partial x_j} =  \boldsymbol{S},  
\end{equation}
where the conservative variables vector $\boldsymbol{U}$, associated fluxes vector $\boldsymbol{F}_j$ and source terms $\boldsymbol{S}$ are defined as below 
\begin{equation} 
\label {eq:2.22}
\boldsymbol{U}=
\begin{bmatrix}
\rho \\
\rho v_i\\
B_i \\
E \\
\psi \\
\end{bmatrix}
, \ \ 
\boldsymbol{F}_j=\begin{bmatrix}
\rho v_j\\
\rho v_iv_j +P_t \delta_{ij} - B_iB_j\\
v_jB_i - B_jv_i + \psi \delta_{ij} \\
(E+P_t)v_j-(\boldsymbol{B}.\boldsymbol{v})B_j\\
c_h^2B_j \\
\end{bmatrix}
, \ \  
\boldsymbol{S}=\begin{bmatrix}
0 \\
0 \\
0 \\
0 \\
-c_h^2/c_p^2\psi \\ 
\end{bmatrix}.
\end{equation} 
Where $i = x, y, z$ stands the different components, $\delta_{ij}$ is the Kronecker symbol and $c_h \in (0, \infty)$ and $c_p \in (0, \infty) $. The additional equation of unphysical variable $\psi$ implies that divergence errors are propagated to the domain boundaries at finite speed $c_h$ and damped at a rate given by $c_h/c_p$. Note that the only source term occurs in the equation for the unphysical variable $\psi$ through the mixed hyperbolic/parabolic correction. 

For the hyperbolic system of conservation laws Eq.~(\ref{eq:2.21}) (without source term), $\boldsymbol{U} \equiv \boldsymbol{U}(\boldsymbol{x}, t): \Re^m \times \Re \to \Re^n $ and $\boldsymbol{F}(\boldsymbol{U}): \Re^n \to \Re^n $. Under the self-similar transformation for 2D $\boldsymbol{\xi}(\xi, \zeta) \equiv \boldsymbol{x}(x, y)/t$~\cite{samtaney1997}, we eliminate the independent variable time $t$ and transform the system that depends only upon the self-similar coordinates $\boldsymbol{\tilde U}(\boldsymbol{\xi}) \equiv \boldsymbol{U} (\boldsymbol{x}, t): \Re^m \to \Re^m, m=2 $. The Eq.~(\ref{eq:2.21}) system becomes 
\begin{equation} 
\label {eq:2.23}
m \bm{\tilde U}  +   \frac{\partial \tilde F_j }{\partial \xi_j}  = \bm{S},          
\end{equation} 
where the self-similar flux vector is defined as $\bm{F}_j = \bm{F}_j - \xi_j \bm{U} $. We add a source term of the same form $-c_h^2/c_p^2\psi $ to the self-similar system in an {\it ad hoc} fashion because adding this source term does not change the nature of physics, and inclusion of this source term allows for additional control of the divergence errors.
The unsteady IVP Eq.~(\ref{eq:2.21}) is thus transformed into the steady BVP Eq.~(\ref{eq:2.23}). The self-similar solution to the BVP Eq.~(\ref{eq:2.23}) is solved using an iterative method, and implemented using the p4est adaptive mesh refinement (AMR) framework~\cite{Burstedde2011,Isaac2015}. Existing Riemann solvers (e.g., Roe, HLLD etc.) can be modified in a relatively straightforward manner and used in the present method. The details of the numerical scheme are presented in Chen \& Samtaney~\cite{Chen2021}.
 
\section{\label{sec:3} Results: Regular Refraction Parallel Field Cases}
In this section we present numerical and analytical solutions for regular refraction of a fast MHD shock at an inclination density interface for the case of a magnetic field that is initially oriented perpendicular to the shock propagation direction.  
Two reference cases, denoted as \boldsymbol{$R_1$} and \boldsymbol{$R_2$}, are discussed in detail below. The reference case \boldsymbol{$R_1$} is characterized by $M = 2, \gamma = 1.4, \eta = 3, \alpha = \pi/4, \beta = 2$~\cite{Samtaney2003}, wherein the magnetic field ($B_y$ is present) is initially perpendicular to the motion of incident shock (aligned $x$ direction). The reference case \boldsymbol{$R_2$} is the same as \boldsymbol{$R_1$} except that $\alpha = \pi/6$. After the discussion of reference case \boldsymbol{$R_1$} we examine the evolution of the wave structures for decreasing $\beta$, i.e., by increasing the strength of the initially applied magnetic field. Thereafter, we examine the effect of changing the inclination angle $\alpha$ by first examining the reference case \boldsymbol{$R_2$} in detail. 
The discussion on irregular refraction in the presence of perpendicular and parallel magnetic fields is deferred until  Section~\ref{sec:4}.

\subsection{\label{subsec:3.1}Reference case \boldmath{$R_1$}}  
\begin{figure}[htbp]   
\centering
\subfigure[IS]{
\includegraphics[scale=0.2]{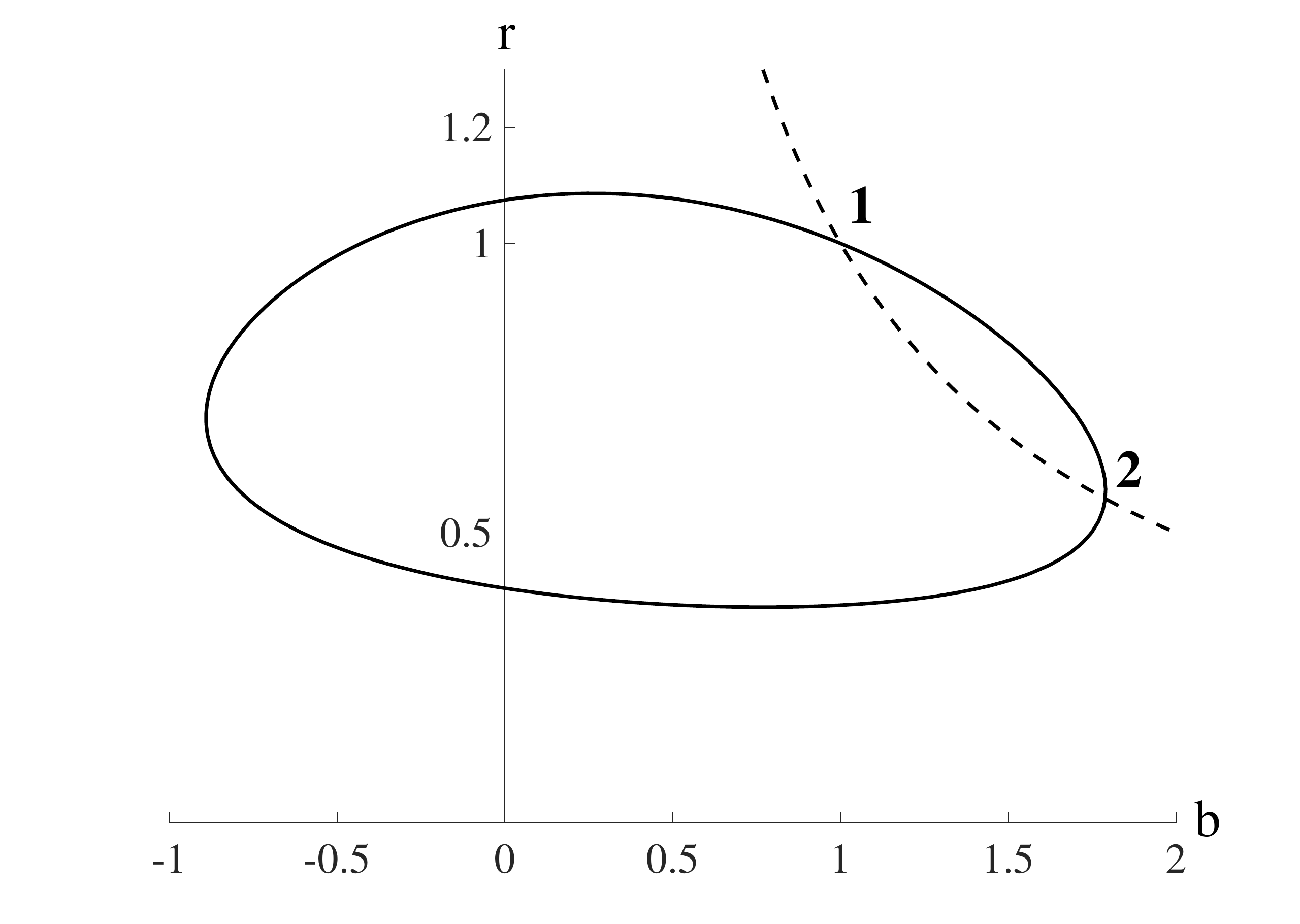}\label{fig:3.1a}
} 
\quad
\subfigure[RF]{ 
\includegraphics[scale=0.2]{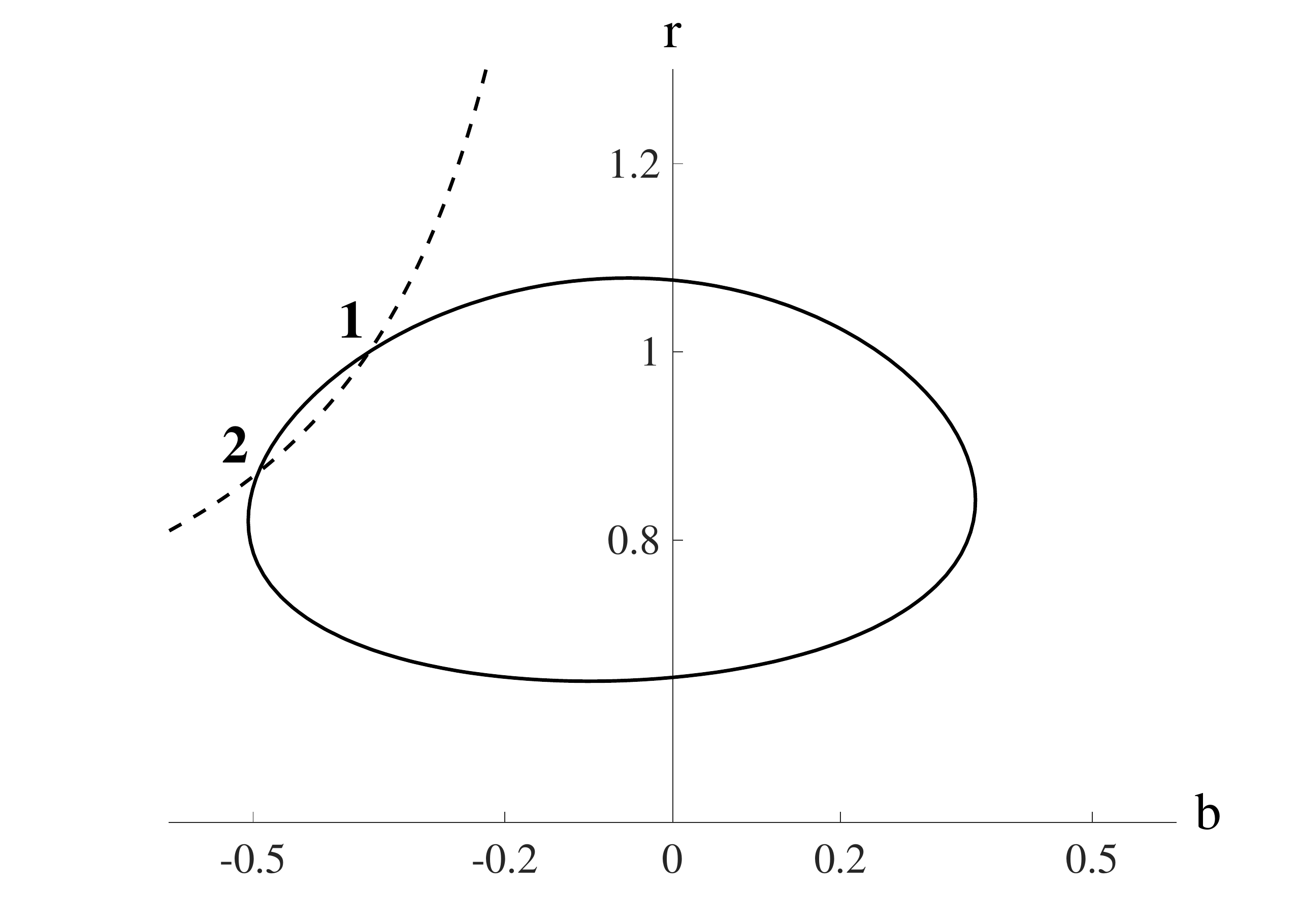}\label{fig:3.1b}
} 
\quad
\subfigure[TF]{ 
\includegraphics[scale=0.2]{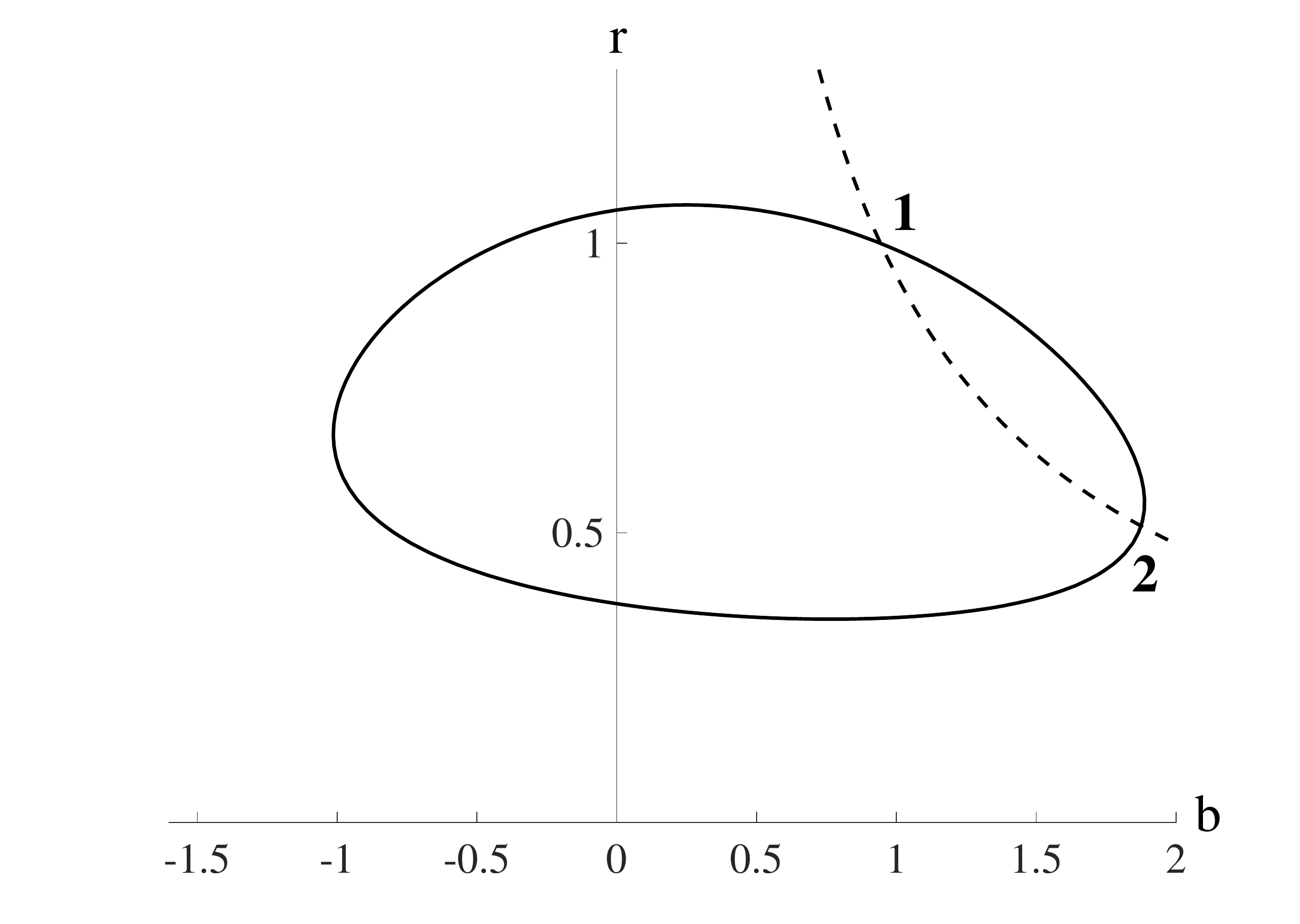}\label{fig:3.1c}
} 
\quad
\subfigure[RS]{ 
\includegraphics[scale=0.2]{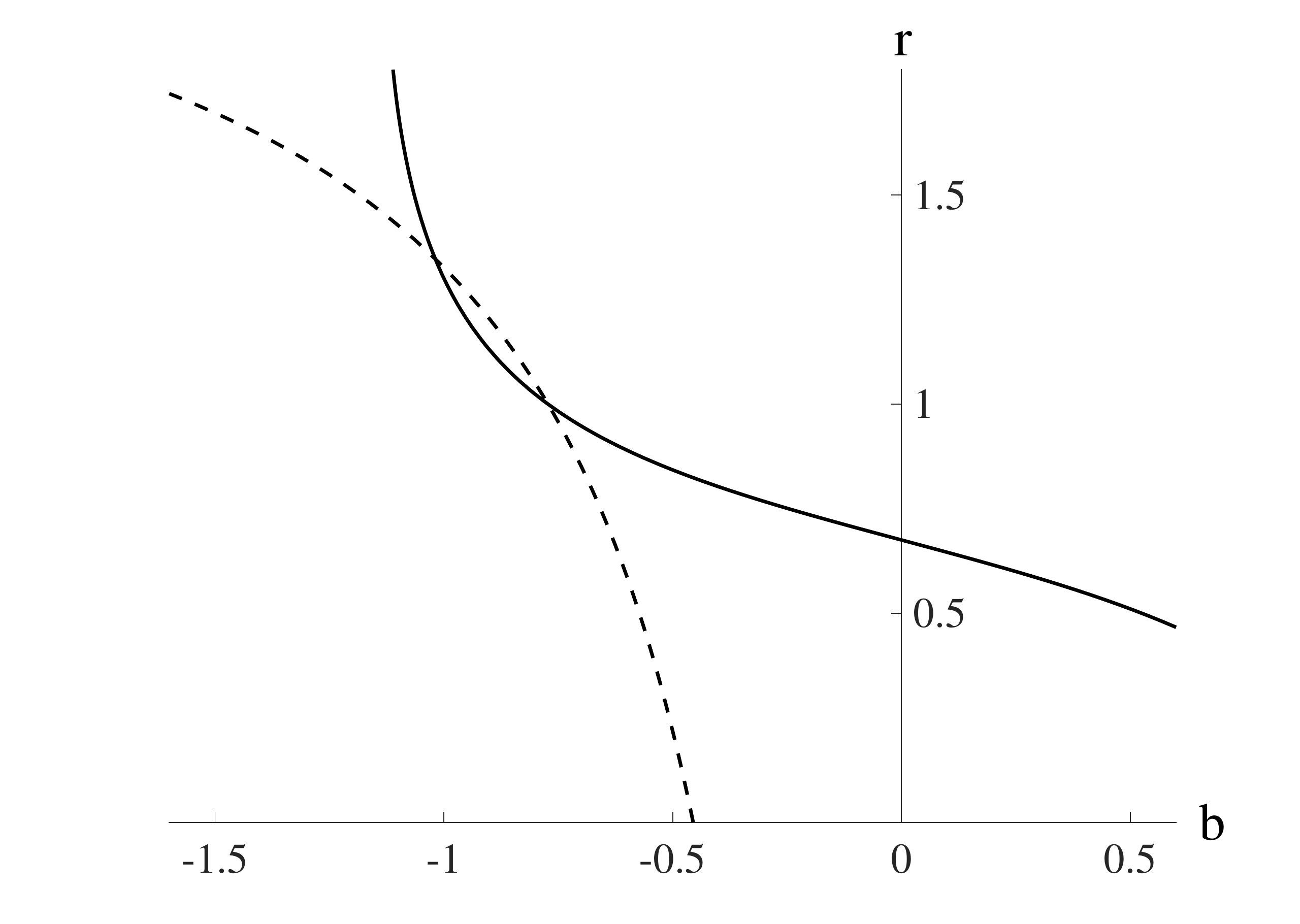}\label{fig:3.1d}
} 
\quad
\subfigure[TS]{ 
\includegraphics[scale=0.2]{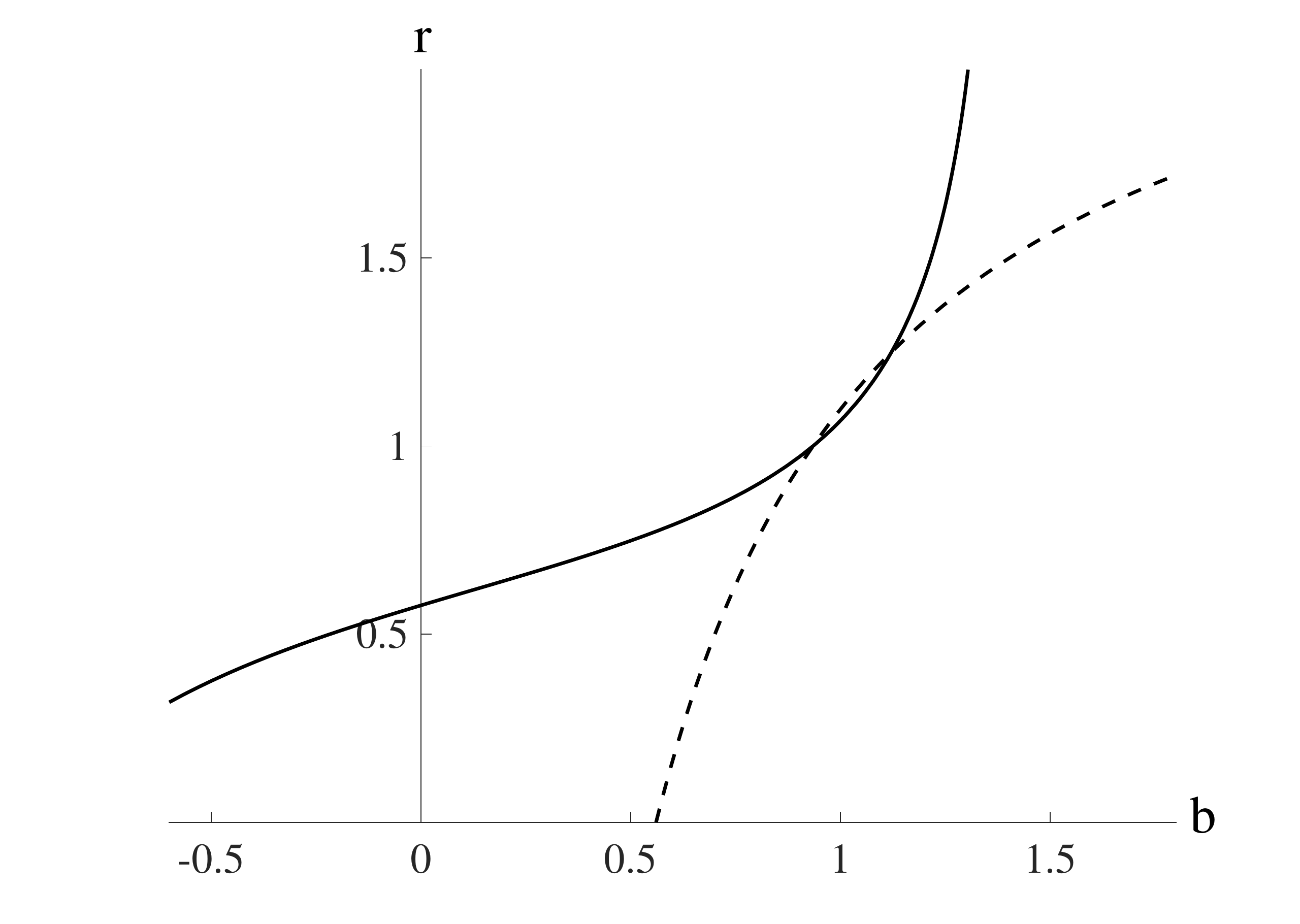}\label{fig:3.1e}
} 
\caption{Graphical solutions of the MHD Rankine-Hugoniot relations for conditions upstream of 5 waves in case \boldsymbol{$R_1$}. Solid (dashed) line denotes $\mathcal F = 0$ $(Z =0)$. }
\label{fig:3.11}       
\end{figure}

We now examine the solution to the reference case \boldsymbol{$R_1$} ($M = 2, \gamma = 1.4, \eta = 3, \alpha = \pi/4, \beta = 2$) where the magnetic field ($B_y$ is present) is initially perpendicular to the motion of incident shock. 
Fig.~\ref{fig:3.1a} shows the graphical solution of the RH relations for the conditions upstream of incident shock. With the exception of $r=1$ (corresponding to the upstream state), there is only one real root $r_0 = 0.559$ which corresponds the downstream IS (state 1). It shows that the incident shock is a fast magnetosonic MHD shock instead of pure hydrodynamic shock.
Discarding $r=1$, the unique real root $r_1$ is $0.872$ for $RF$, while it is $0.512$ for $TF$, see in Fig.~\ref{fig:3.1b} and Fig.~\ref{fig:3.1c}, respectively. It indicates that $RF$ and $TF$ are also fast mode MHD shocks. 
However,  Fig.~\ref{fig:3.1d} shows the unique real root $r_2>1$ (again discarding $r=1$), which implies $RS$ should be either an expansion fan or a compound wave. 
Additionally, the sign of magnetic field does not change over this wave, indicating that the $RS$ is an expansion fan through which  only the magnitude of magnetic field changes. 
The same observation for $TS$ is found in Fig.~\ref{fig:3.1e}; hence we consider $TS$ also to be an expansion fan. 
It demonstrates that the solution includes three fast shocks and two slow-mode expansion fans. 
The four waves $RF, RS, TF$ and $TS$ are found to lie at $\phi_1 = 0.370255, \phi_2 = 0.765571, \phi_3 = 1.232414$ and $\phi_4 = 1.044445$, respectively. We also analytically compute the angular width of expansion fan $\Delta \phi_{ERS} = 0.330408$ for $RS$, whereas $\Delta \phi_{ETS} =0.180417$ for $TS$, respectively. 

\begin{figure}[htbp]   
\centering
\subfigure[]{
\includegraphics[scale=0.17]{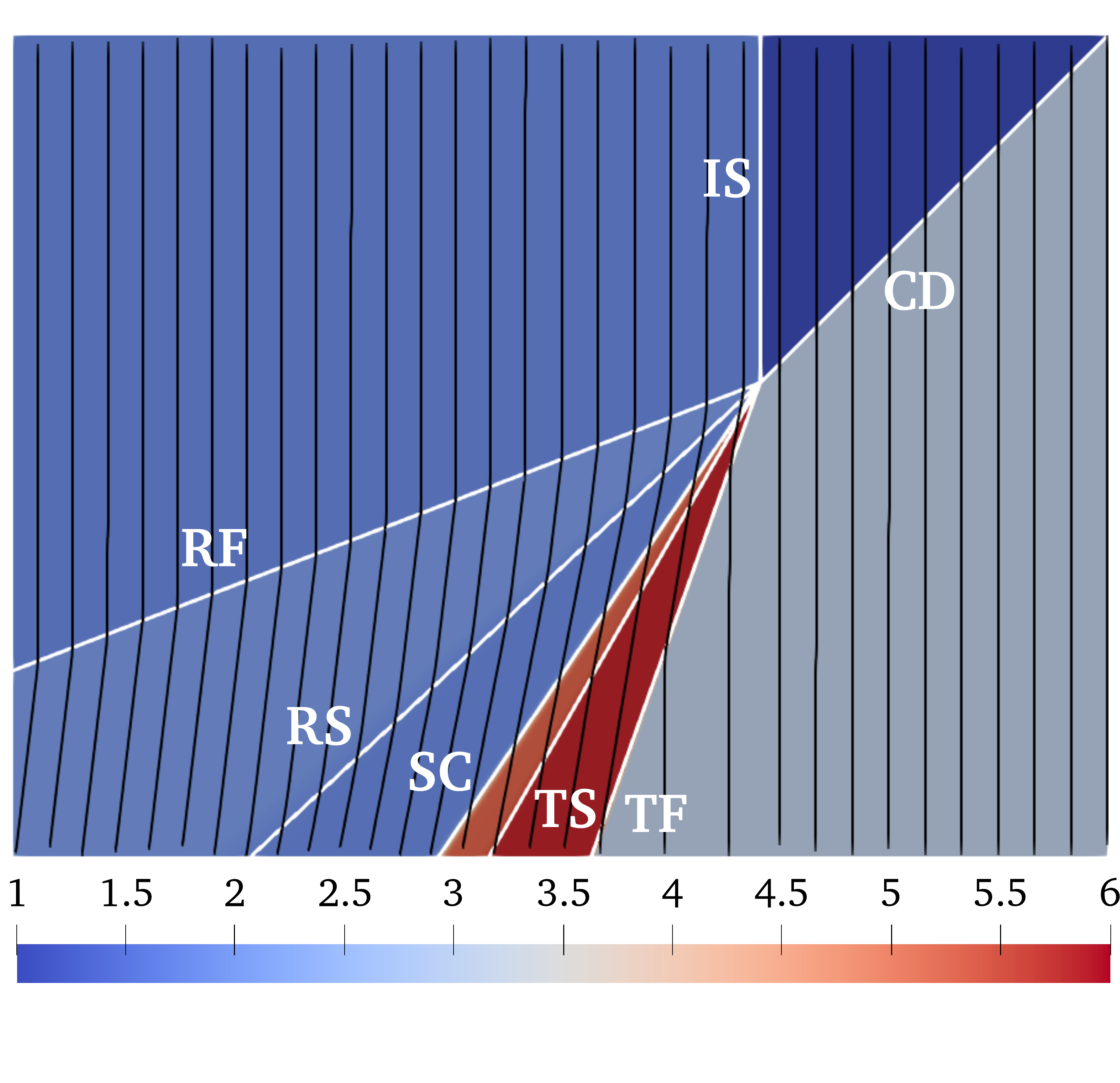}\label{fig:3.12a}
} 
\subfigure[]{ 
\includegraphics[scale=0.17]{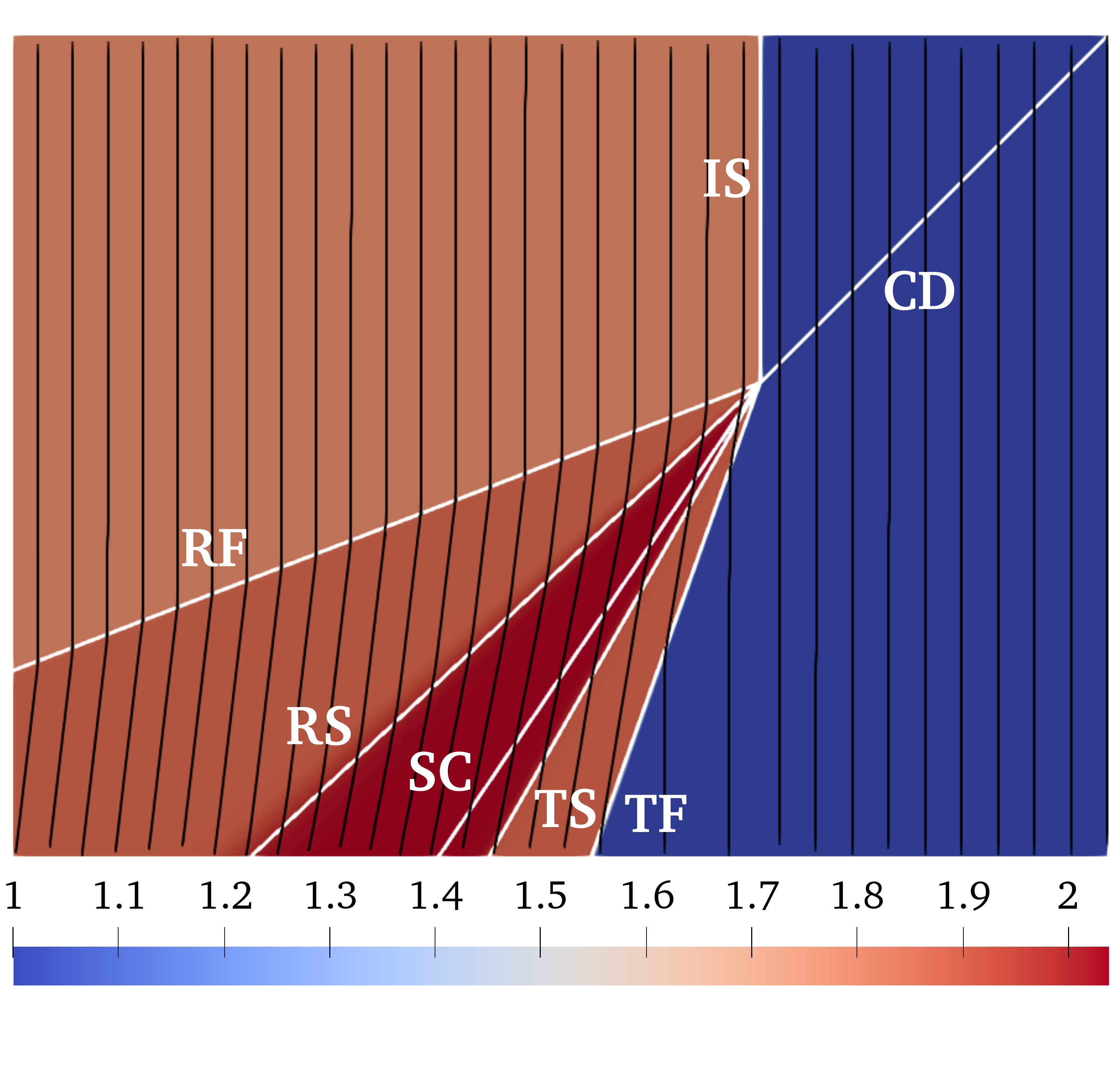}\label{fig:3.12b}
} 
\subfigure[]{ 
\includegraphics[scale=0.17]{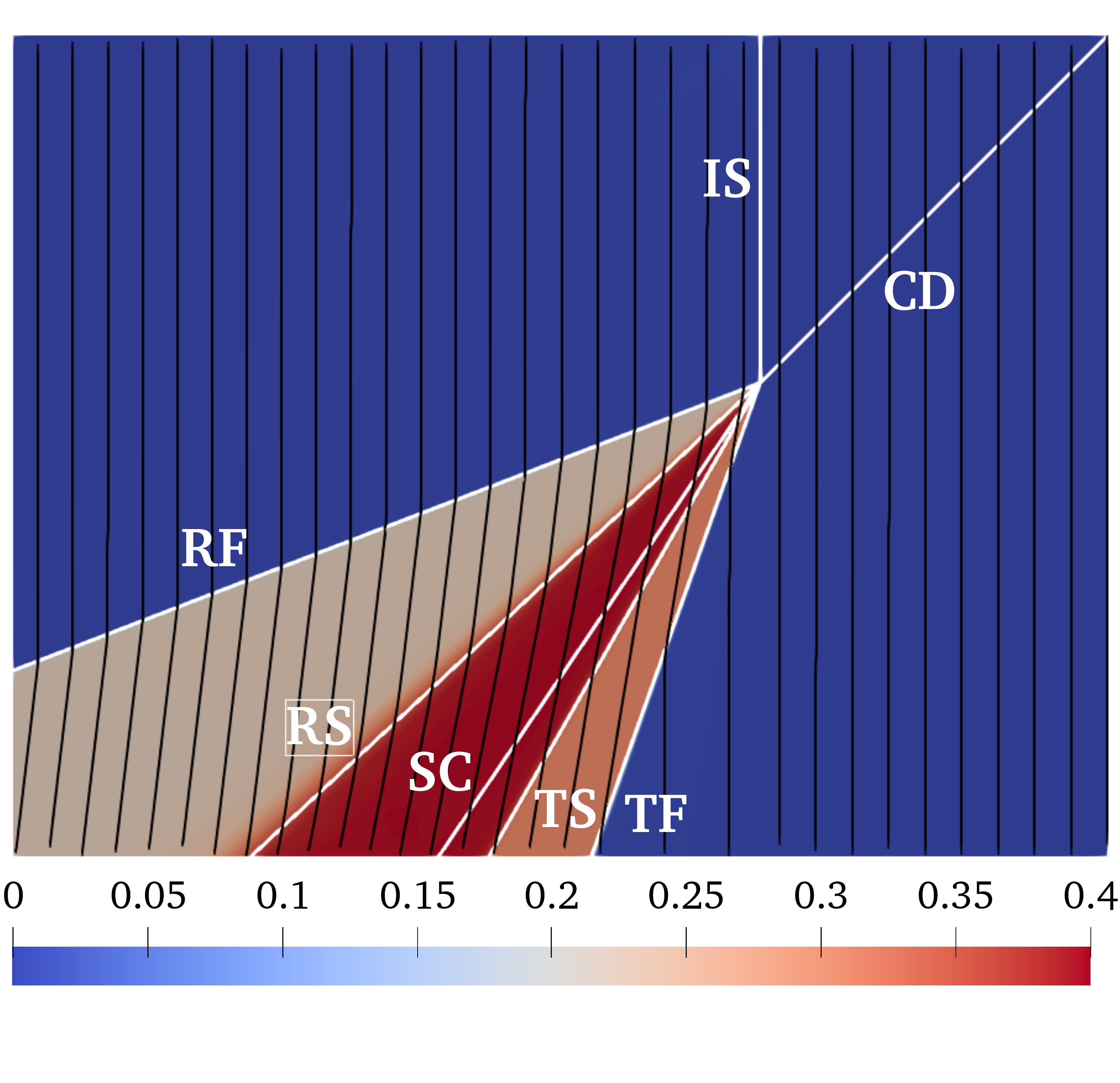}\label{fig:3.12c} 
} 
\caption{Analytical wave structure for the case \boldsymbol{$R_1$} overlaid on the numerical results displayed by density (a), $B_y$ (b) and $B_x$ (c) fields, respectively. Magnetic field lines are shown by black solid lines. }
\label{fig:3.12}       
\end{figure}

\begin{figure}[htbp]   
\centering
\subfigure[]{
\includegraphics[scale=0.44]{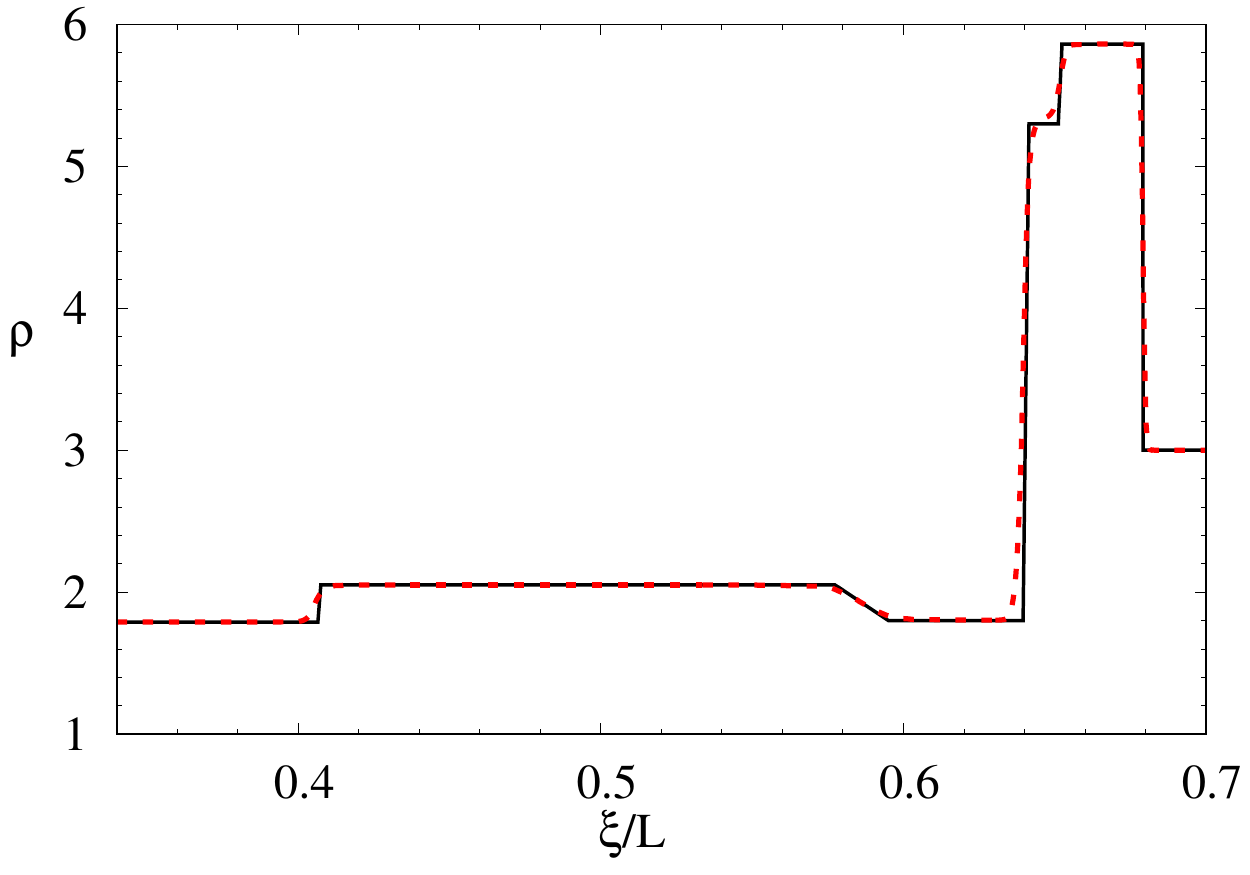}\label{fig:3.13a}
} 
\subfigure[]{ 
\includegraphics[scale=0.44]{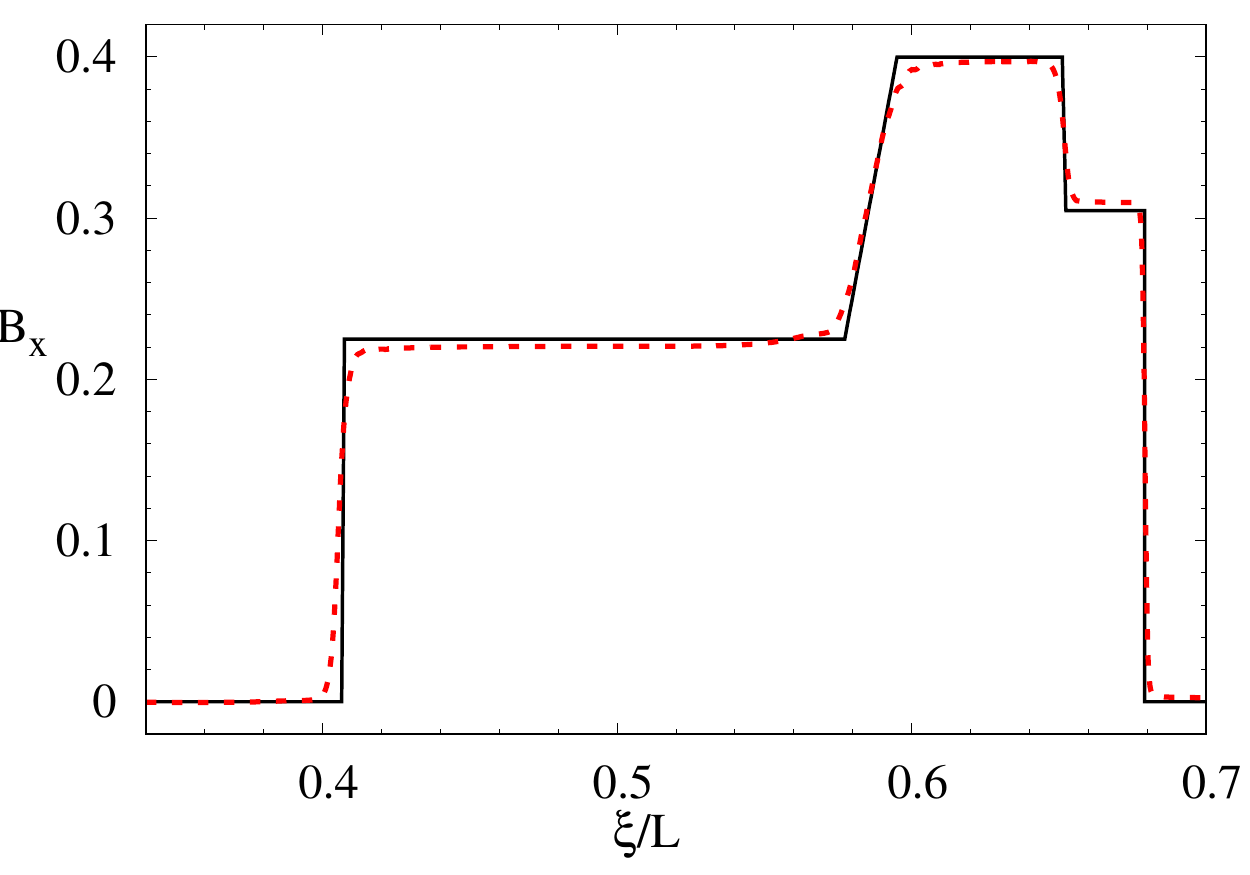}\label{fig:3.13b}
} 
\subfigure[]{ 
\includegraphics[scale=0.44]{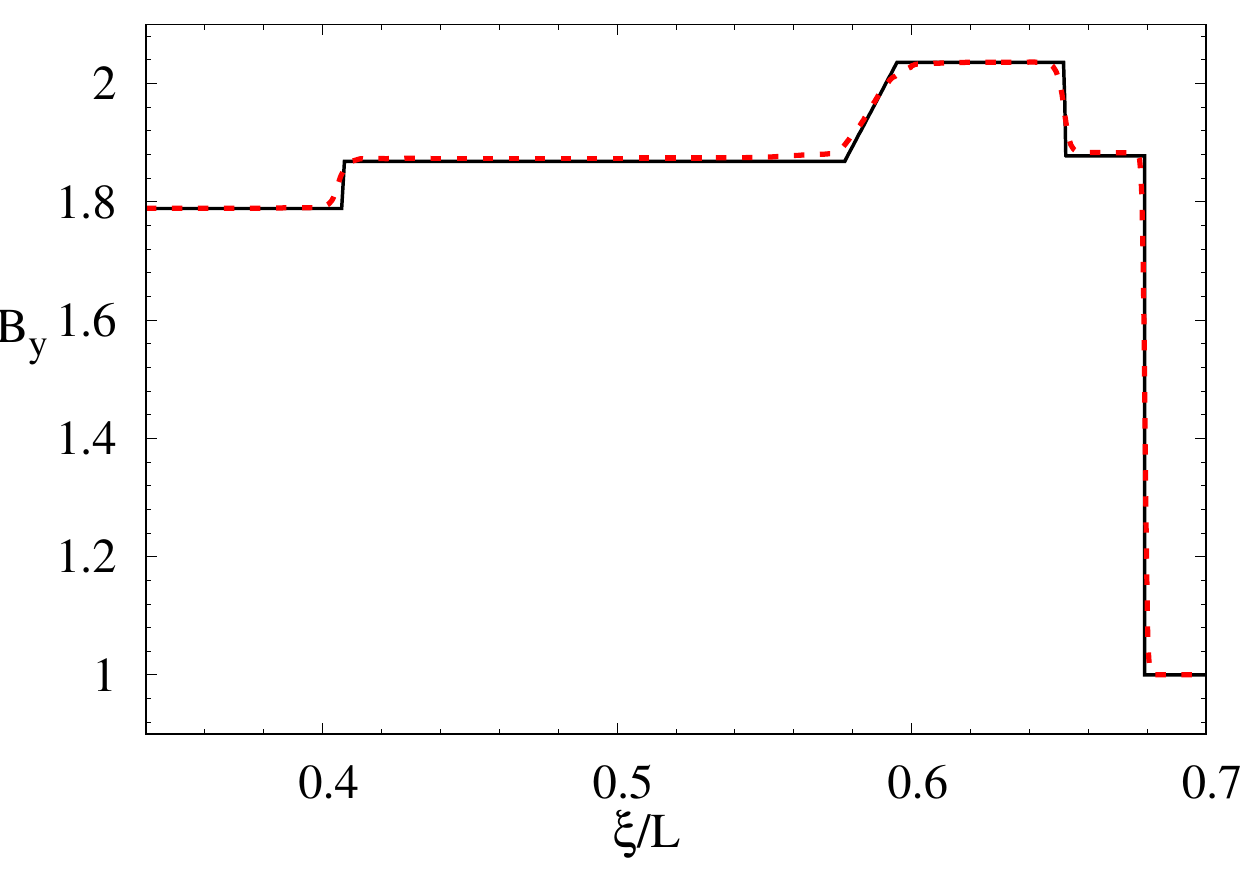}\label{fig:3.13c}
} 
\caption{Density $\rho$ (a), component $B_x$ (b), and $B_y$ (c) of magnetic field profiles from the numerical results at $\zeta /L = 0.6667$ compared to profiles from analytical solution. Solid (dashed) line denotes analytical (numerical) solution.}
\label{fig:3.13}        
\end{figure}

In Fig.~\ref{fig:3.12}, we overlay the analytical wave structure (the locations of waves) on the numerical results displaying the density and magnetic fields. 
Fig.~\ref{fig:3.12a} shows that the density field clearly displays the location of shocked contact (denoted as $SC$), over which the magnetic field remains continuous. The x-component of the magnetic field, $B_x$, clearly displays the locations of the weaker $RF$ and $RS$ waves that have small density jumps associated with them and these waves are not so clearly discerned in the density plot, see in Fig.~\ref{fig:3.12c}. 
Both $RS$ and $TS$  increase the magnitude of magnetic field without changing its sign but it is not obvious from the 2D field plots of density and magnetic fields that $RS$ and $TS$ are expansion fans. To show that both are indeed expansion fans, we plot $\rho, B_x$ and $B_y$ profiles along a horizontal line at $\zeta/L = 0.6667$ in Fig.~\ref{fig:3.13}. 
From left to right in Fig.~\ref{fig:3.13a}, the waves are: fast-mode shock $RF$, slow-mode expansion fan $RS$, shocked contact $SC$, slow-mode expansion fan $TS$ and fast-mode shock $TF$, respectively. The first two waves propagate to left, the remaining three waves move to the right. 
We also note that the magnetic field does not change its sign passing through all waves, and hence there are neither intermediate shocks nor compound waves for the strongly planar solution in the reference case \boldsymbol{$R_1$}. It implies that the planar solution is identical to the strongly planar solution for the case \boldsymbol{$R_1$}.   
It is interesting to contrast this with the case of a parallel magnetic field~\cite{Samtaney2003,Wheatley2005JFM}, where  it was shown that $TS$ is $2 \to 4$ intermediate shock and $RS$ is slow shock for strongly planar solution, whereas $TS$ is replaced by a rotational discontinuity followed downstream by a slow shock for the planar solution. 
With the specific parameters of case \boldsymbol{$R_1$}, the wave structure resulting from perpendicular magnetic field is unique, while the wave structure is not unique for the parallel magnetic field case which exhibits differences between planar and strongly planar situations.   
We further note,  from Fig.~\ref{fig:3.12} and Fig.~\ref{fig:3.13},  there is close agreement between the analytical solution and the numerical results insofar the wave structures is concerned. 
%
\subsection{\label{subsec:3.2}Evolution of wave structure with \boldmath{$\beta$} }
We now examine how solutions to the regular MHD shock refraction problem vary as magnetic field strength (characterized by $\beta$) is varied. 
The problem is characterized with the same parameters in the case \boldsymbol{$R_1$} except $\beta \in (0.5, 10^6)$.
The solutions for strong magnetic fields $\beta \in (0.5, 10)$ are presented in Section~\ref{subsec:3.2.1}, while the solutions for large $\beta$ are shown in Section~\ref{subsec:3.2.2}. 
\subsubsection{\label{subsec:3.2.1}Solution behavior for $\beta \in (0.5, 10)$ }
\begin{figure}[h]   
\centering
\subfigure[]{
\includegraphics[scale=0.6]{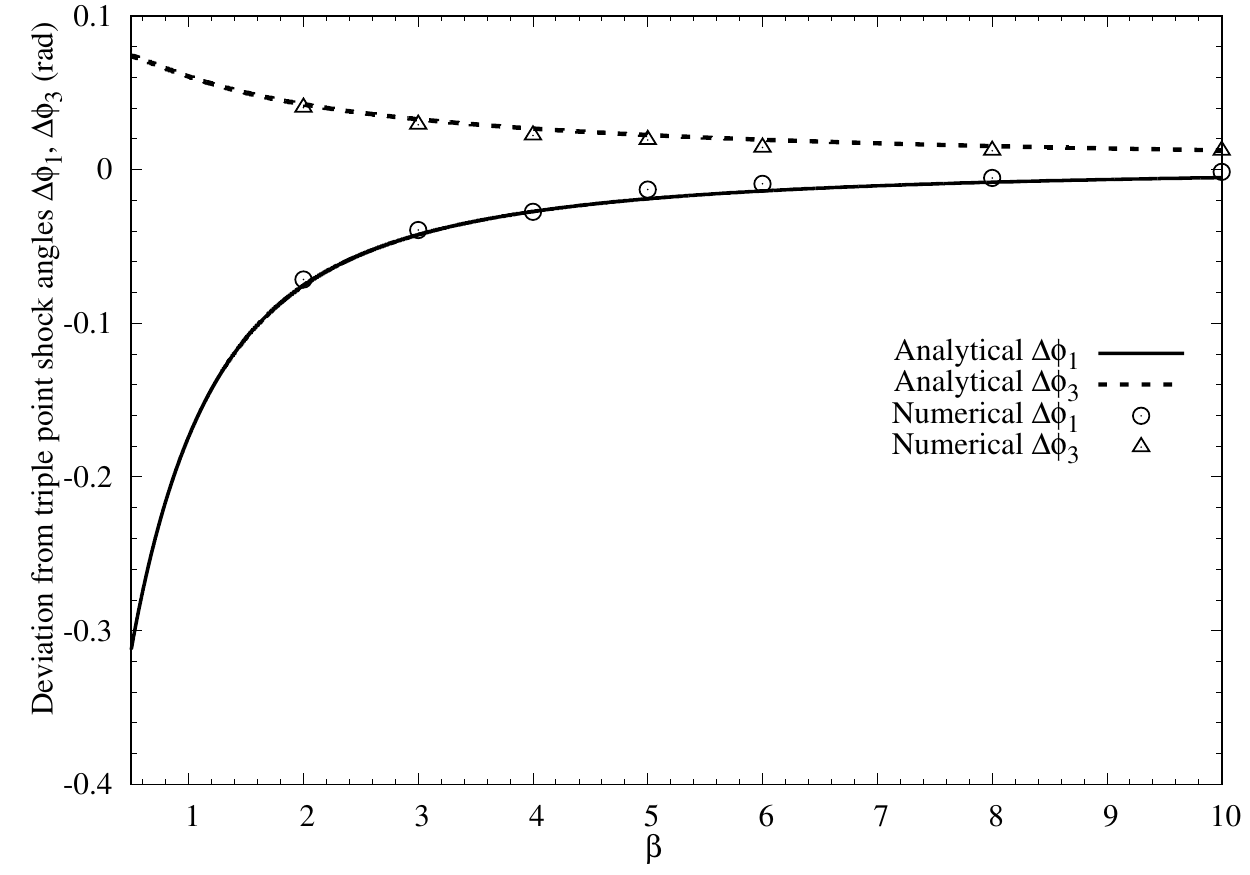}\label{fig:3.21a}
} 
\subfigure[]{ 
\includegraphics[scale=0.6]{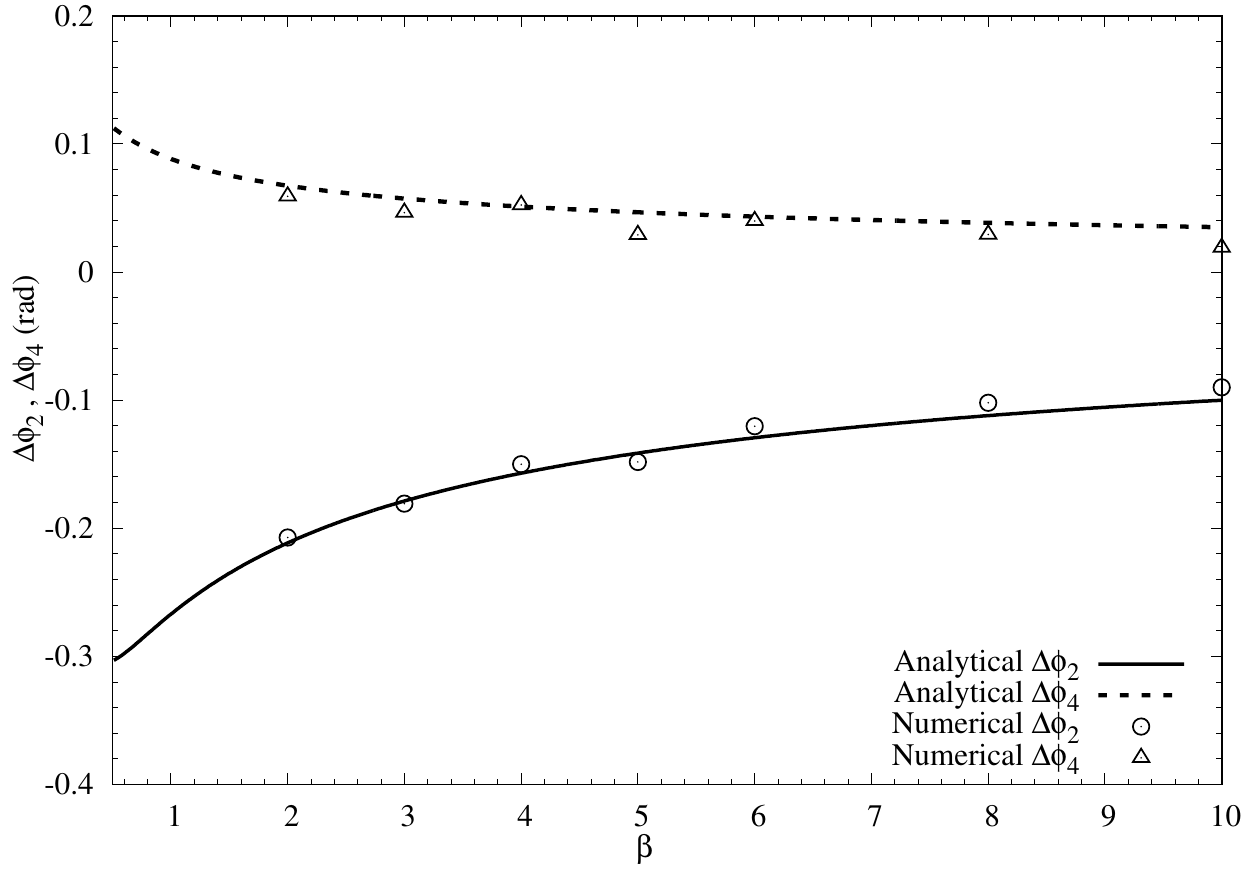}\label{fig:3.21b}
} 
\caption{(a) Deviation of the fast shock angles from their corresponding values in the hydrodynamic triple-point, $\Delta \phi_1$ and $\Delta \phi_3$ for $RF$ and $TF$, respectively; (b) deviation of slow-mode expansion fans angles $\Delta \phi_2$ and $\Delta \phi_4$ for $RS$ and $TS$. For hydrodynamic case, $\phi_{1,hydro} = 0.4454$, and $\phi_{3,hydro} =  1.2771$. }
\label{fig:3.21}       
\end{figure}

In hydrodynamic case, regular refraction leads to a triple point, i.e., there are three shocks (incident, reflected and transmitted) which meet at a single point. 
The angles of shocks $R$ and $T$ in hydrodynamic triple-point solution to the shock refraction problem are computed with $\beta^{-1} = 0$, $M= 2, \alpha = \pi/4, \gamma = 1.4$ and $\eta = 3$. We plot the deviation of the angles of the fast-mode shocks from their corresponding triple point values in figure~\ref{fig:3.21a}.  
Here, $\Delta \phi_1$ corresponds to the angle of $RF$ minus the angle of $R$, while $\Delta \phi_3$ corresponds to the angle of $T$ minus the angle of $TF$, respectively. 
As the field strength is decreased, or as $\beta$ is increased, the angles of the fast-mode shock in MHD tend toward the triple-point values from hydrodynamics. 
For $\beta < 3$ the angle deviation from the triple point values increases as $\beta$ is decreased, the change of $RF$ is stronger than that of $TF$. 
Beyond $\beta=3$, the evolution of angle deviation changes more gradually and tends to be linear at larger $\beta$ values. 
Meanwhile, the location angles $\phi_2$ (for $RS$) and $\phi_4$ (for $TS$) tend toward the $SC$ with increasing of $\beta$, as shown in Fig.~\ref{fig:3.21b}. 
The $\Delta \phi_2$ ($\Delta \phi_4$) corresponds the difference between $\phi_2$ ($\phi_4$) and angular location of $SC$.
The angular difference $\mid \Delta \phi_2 \mid$ decreases from $0.3$ to around $0.1$ with increasing of $\beta$, while $\Delta \phi_4$ decreases more slightly from $0.11$ to $0.04$. 
These variations imply that the reflected waves are more strongly affected than the transmitted waves by the varying strength of magnetic fields. 
In addition, we note that there is a close agreement between the analytical solutions and the numerical results for these strongly planar shock refractions. 
Note that there is no transition of wave configuration with increasing of $\beta$, and the solution consists of three fast shocks and two slow-mode expansion fans.
In contrast,  for the case with the presence of parallel magnetic fields, as $\beta$ increased the $RS$ and $TS$ transition from slow shocks to $2\to4$ intermediate shocks, and then become slow compound waves. Such transitions were noted by Wheatley \etal ~\cite{Wheatley2005JFM} for the parallel field case.
\begin{figure}[h]   
\centering
\subfigure[]{
\includegraphics[scale=0.6]{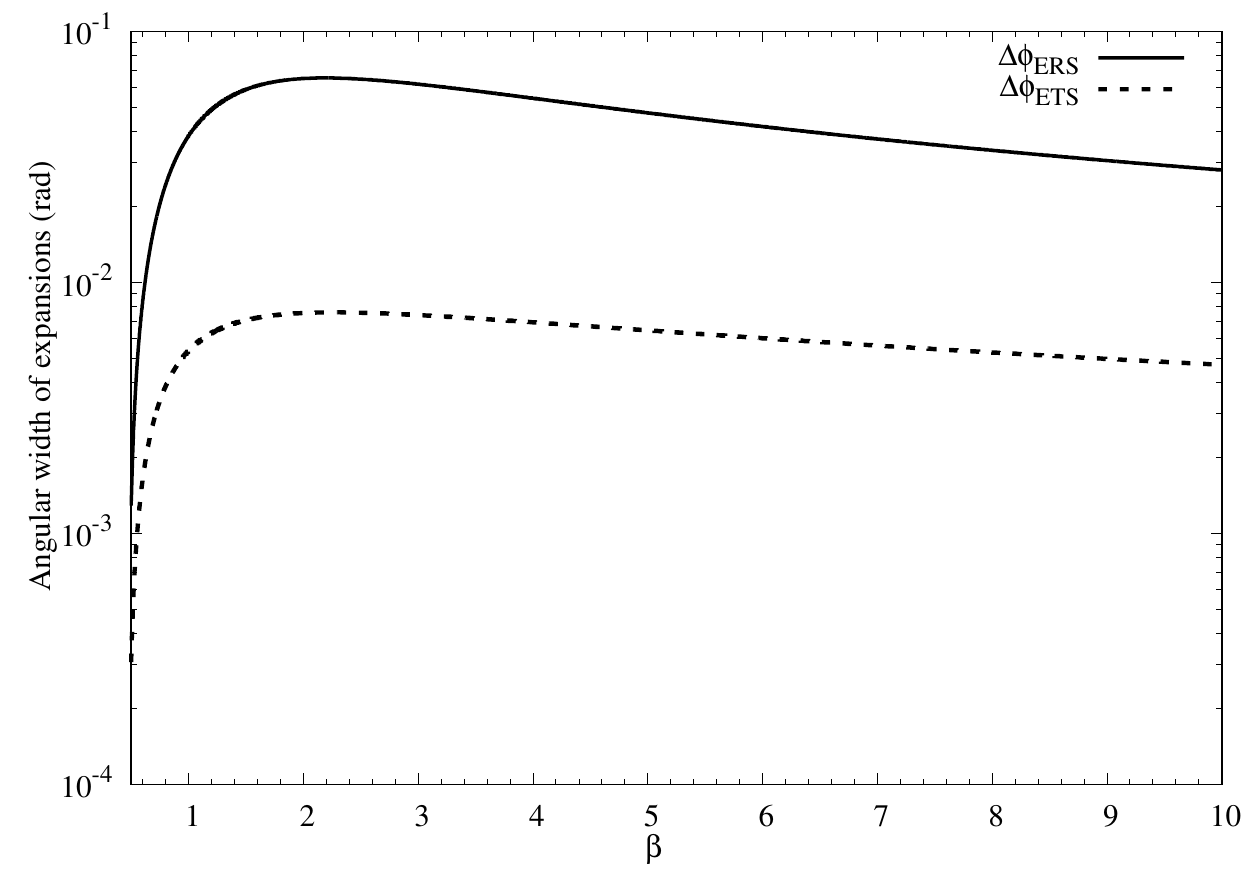}\label{fig:3.22a}
} 
\subfigure[]{ 
\includegraphics[scale=0.6]{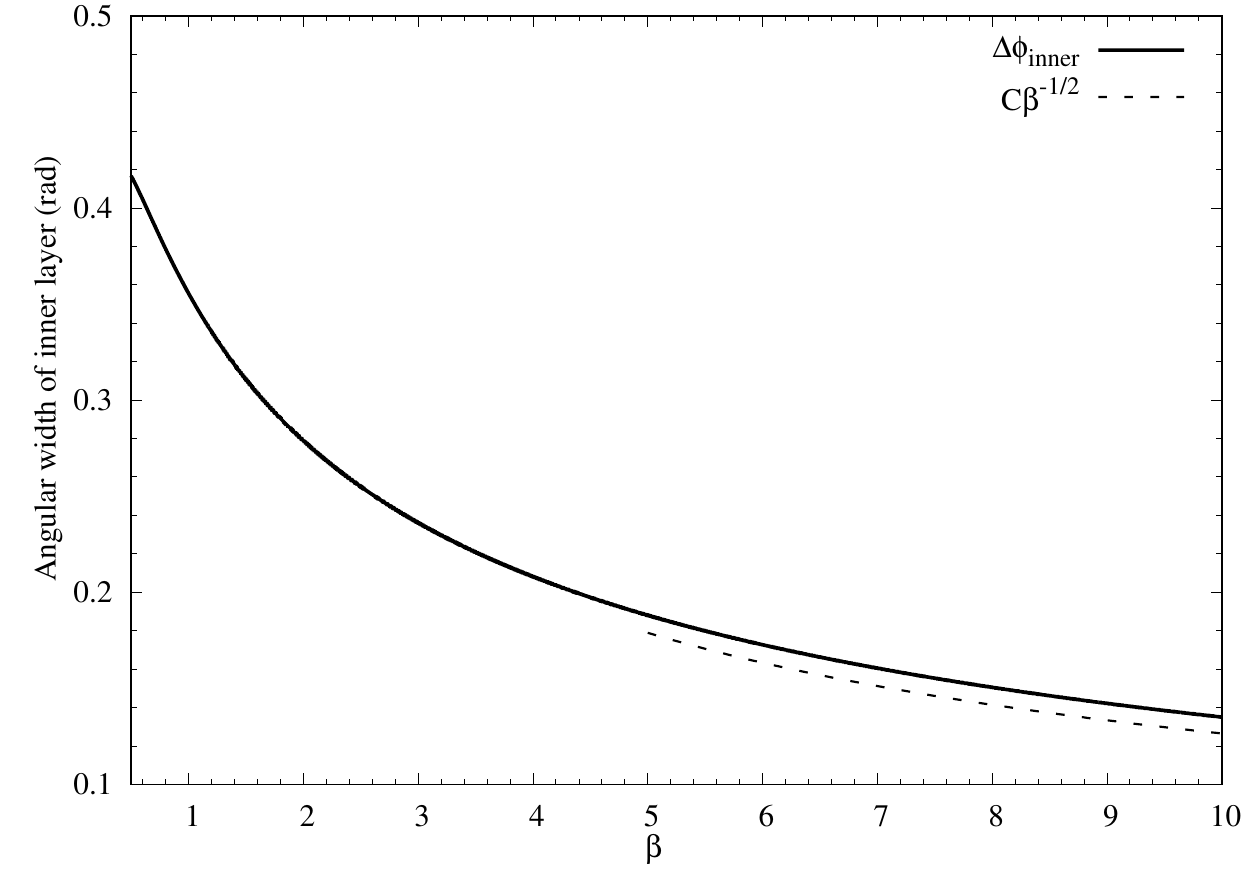}\label{fig:3.22b}
} 
\caption{(a) Angular width of slow-mode expansion fans, $\Delta \phi_{ERS}$ ($\Delta \phi_{ETS}$) denotes the angle width of $RS$ ($TS$) expansion fan; (b) angular width of inner layer. }
\label{fig:3.22}       
\end{figure} 

In Fig.~\ref{fig:3.22a}, we show the variation of the angular width of expansion fans,  $RS$ and $TS$, as $\beta$ is increased.
Here, we plot only the analytical solutions since it is quite challenging to effectively resolve angular widths to the order of $10^{-3}$ from numerical results. 
At $\beta = 0.5$, the angular width for $RS$ is $\Delta \phi_{ERS}=1.344\times 10^{-3}$, whereas it is $\Delta \phi_{ETS} =3.191\times 10^{-4} $ for $TS$. 
When a strong magnetic field is present, the angular width of both expansion fans significantly increases as $\beta$ is increased. The angular width of the reflected expansion reaches a maximum value $\Delta \phi_{ERS} = 6.50\times 10^{-2}$ at $\beta = 2.3627$, while the angular width of the transmitted expansion fan reaches a maximum value of $\Delta \phi_{ETS} = 7.596\times 10^{-3}$  at $\beta = 2.2673$. 
Until $\beta\approx 2.4$ (the location of the maxima), the angular width of the expansion fans is also accompanied by a significant increase in the angular locations of $RS$ and $TS$ (Fig.~\ref{fig:3.21b}). 
The increase in the $RS$ location angle is larger than the than the location angle of $TS$, suggesting that the angular width (defined as the angle from the leading wave in the $RS$ wave group to the leading wave in the $TS$ wave group) of the inner layer decreases as $\beta$ increased (see Fig.~\ref{fig:3.22b}).
After reaching the maximum values of angular width $\Delta \phi_{ERS}$ and $\Delta \phi_{ETS}$, the angular location of $RS (\phi_2)$ and $TS (\phi_4)$ continue to increase with $\beta$. 
It leads to a continuous smooth and monotonic decrease in the angular width of the inner layer with increasing $\beta$, the angular width of the inner layer scales as $\beta^{-1/2}$ for $\beta > 5$. 
To summarize, in the range of strong magnetic fields $\beta \in (0.5, 2.4)$, the wave configurations change significantly and the angular width of expansion fans $\Delta \phi_{ERS}$ and $\Delta \phi_{ETS}$ increase as $\beta$ increased. In the range of moderate magnetic fields, the wave configurations change slightly and the angular width of expansion fans $\Delta \phi_{ERS}$ and $\Delta \phi_{ETS}$ decrease as $\beta$ increased.   
\subsubsection{\label{subsec:3.2.2}Solution behavior at large $\beta$}
For cases with parameters $M= 2, \alpha = \pi/4, \gamma = 1.4, \eta = 3$ and $\beta > 0.5$, the strongly planar solution consists of three fast shocks and two slow-mode expansion fans (Fig.~\ref{fig:3.13}), and there is no transition of wave configuration with increasing of $\beta$. Presently we are concerned with the evolution of wave structure at large $\beta$ up to $10^{6}$. 
We plot only analytical angular widths since it is very challenging to obtain the numerical results that resolve well angular widths of order $10^{-4}$. 

\begin{figure}[h]   
\centering
\subfigure[]{
\includegraphics[scale=0.4]{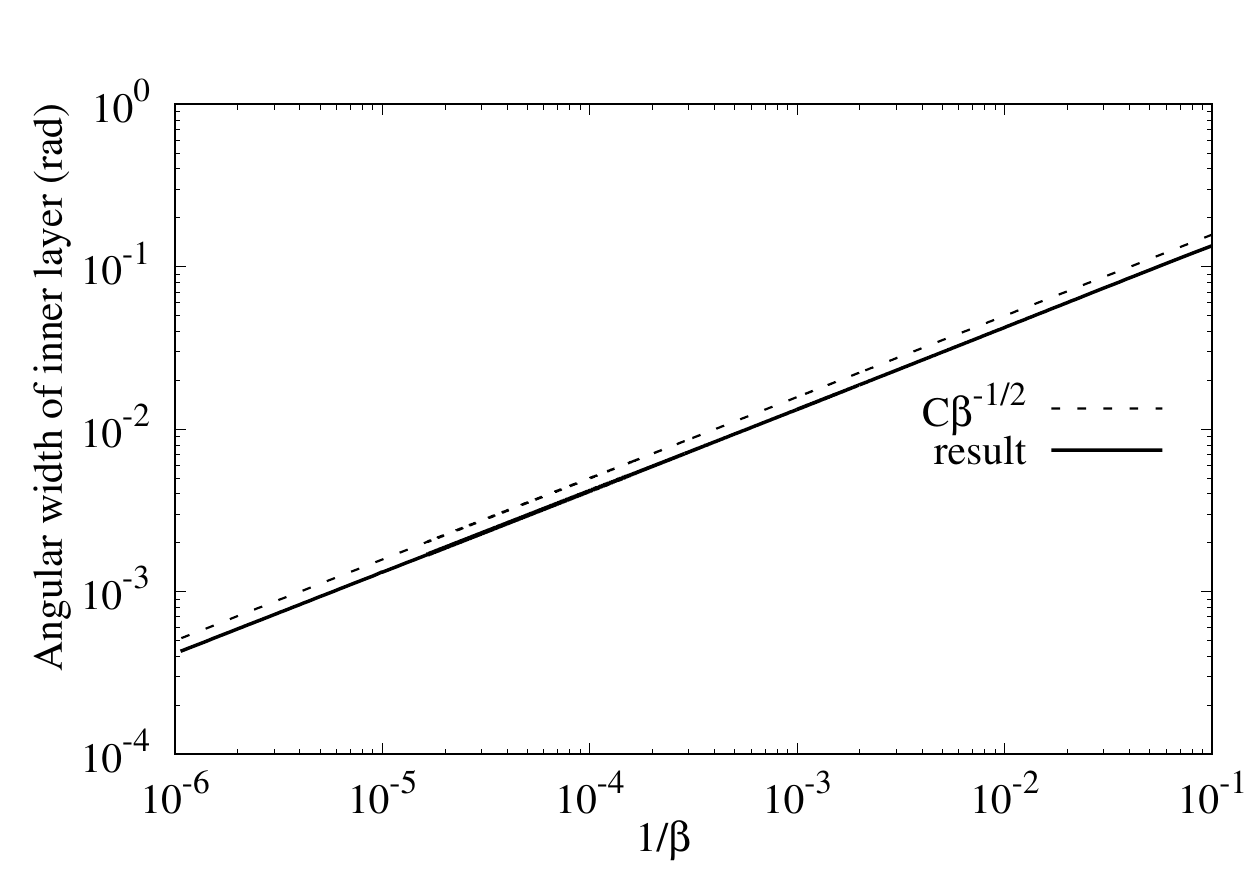}\label{fig:3.23a}
} 
\subfigure[]{ 
\includegraphics[scale=0.4]{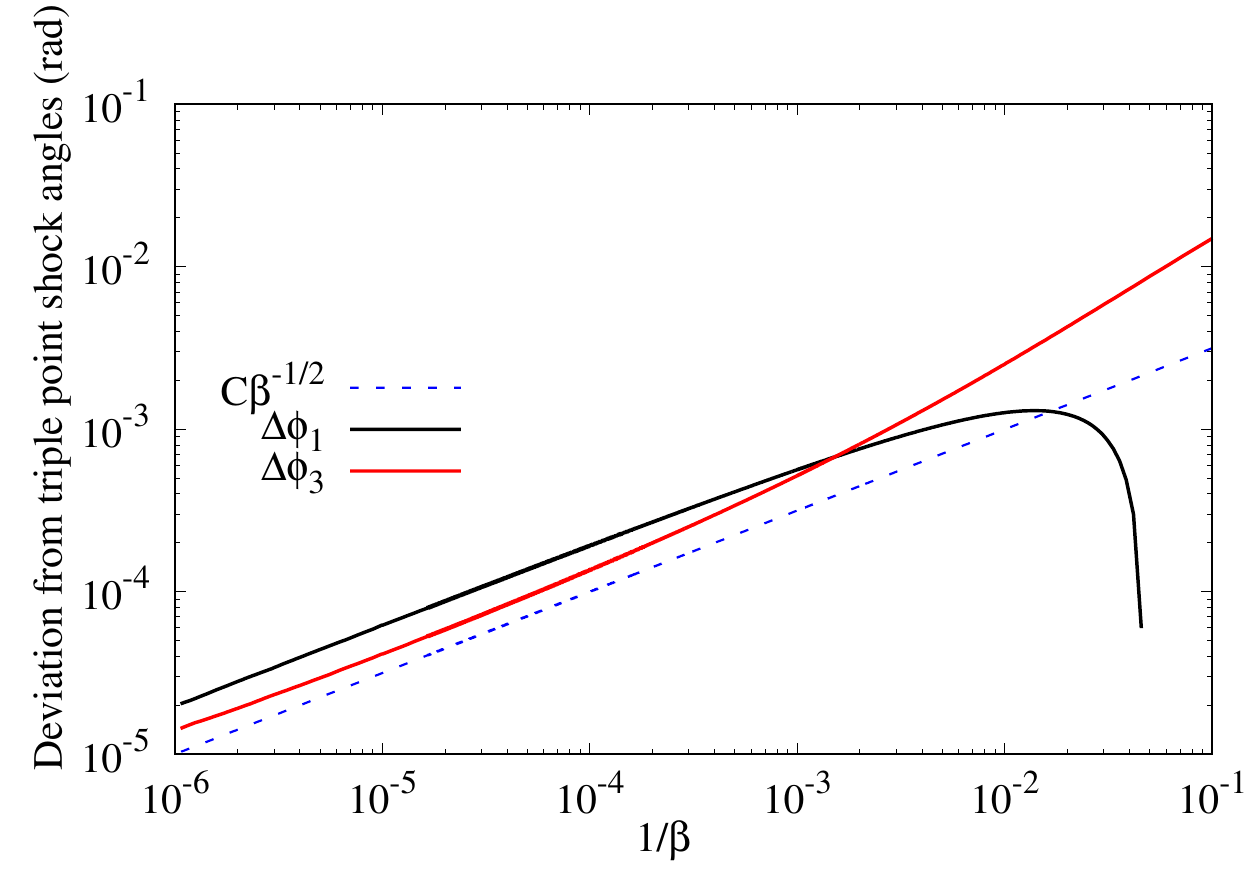}\label{fig:3.23c}
} 
\subfigure[]{ 
\includegraphics[scale=0.4]{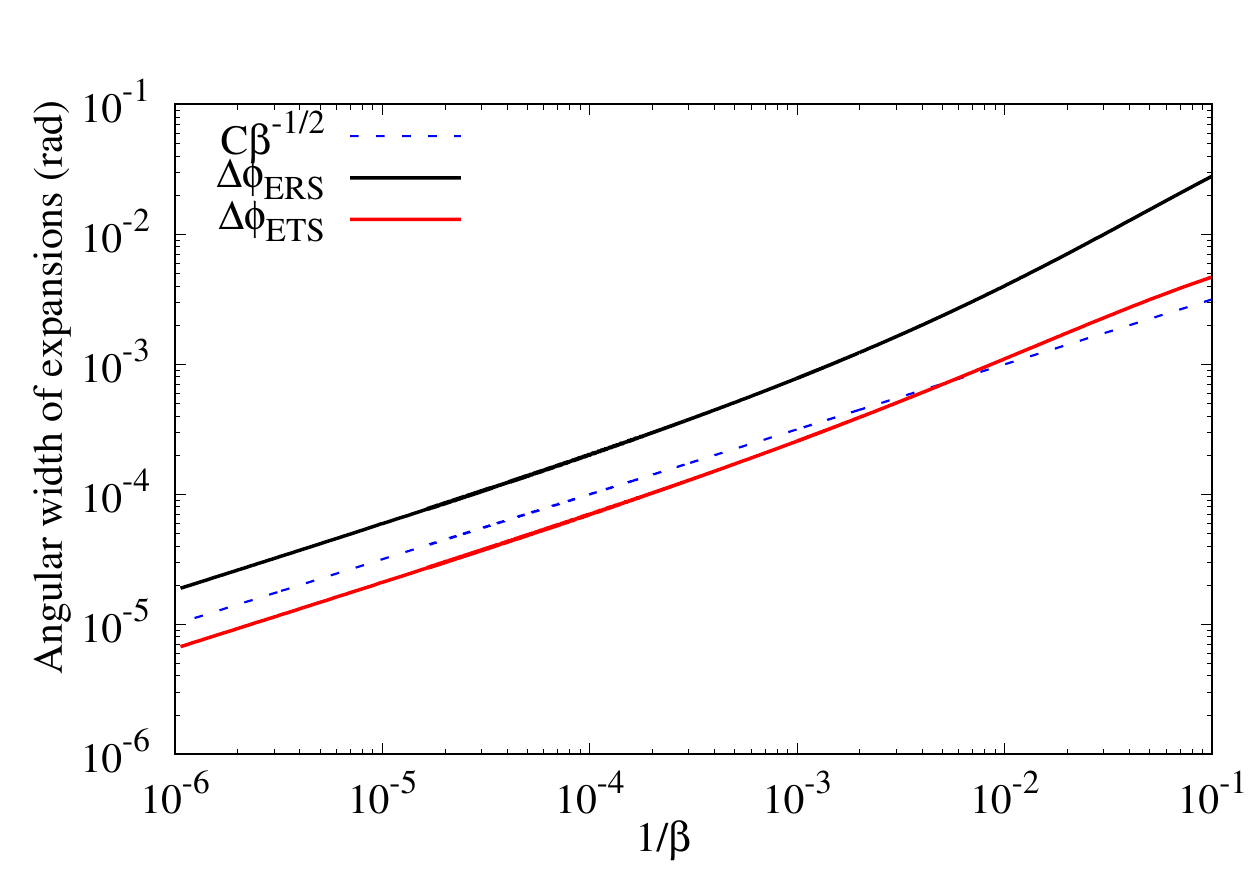}\label{fig:3.23d}
} 
\caption{(a) Angular width of inner layer; (b) deviation of the fast shock angles from their corresponding values in the hydrodynamic triple-point, $\Delta \phi_1$ and $\Delta \phi_3$ for $RF$ and $TF$, respectively; (c) angular width of slow-mode expansion fans, $\Delta \phi_{ERS}$ and $\Delta \phi_{ETS}$ for $RS$ and $TS$, respectively. }
\label{fig:3.23}       
\end{figure}

Fig.~\ref{fig:3.23a} shows that as magnetic field weakens, the angular width of the inner layer diminishes. 
The slope of the angular width of the inner layer versus $\beta^{-1}$, when plotted on a logarithmic scale, reveals that the angular width of the inner layer scales as $\beta^{-1/2}$, i.e., the inner layer angular width is directly proportional to the applied magnetic field magnitude $B$. 
Fig.~\ref{fig:3.23c} shows the deviation of $RF$ and $TF$ from triple-point shock angles versus $\beta^{-1}$. 
It reveals that as magnetic field weakens $\beta^{-1} < 10^{-4}$, the deviation of two fast shocks from triple-point shock angles diminish, and both two scale as $\beta^{-1/2}$. 
This scaling is also found for the evolution of angular width of two expansion fans if $\beta^{-1} < 10^{-4}$ (see Fig.~\ref{fig:3.23d}). 
These observations suggest that, in the limit as $\beta \to \infty$, the location of $RF$ $(TF)$ tends to be identical to $R$ $(T)$ of the corresponding hydrodynamic case. 
The jumps across the inner layer, which are equal to those across the hydrodynamic $SC$ in the limit, are supported by two expansion fans within the layer.
In other words, as $\beta \to \infty$ the MHD solution is identical to the corresponding hydrodynamic triple-point solution, with the exception that the hydrodynamic $SC$ is replaced by the inner layer surrounded by the two slow expansion fans.
This observation is consistent with the conclusion of the case in the presence of parallel magnetic field~\cite{Wheatley2005JFM}. 
%

\subsection{\label{subsec:3.3}Evolution of wave structure with change of $\alpha$ }
\begin{figure}[htbp]   
\centering
\subfigure[]{
\includegraphics[scale=0.2]{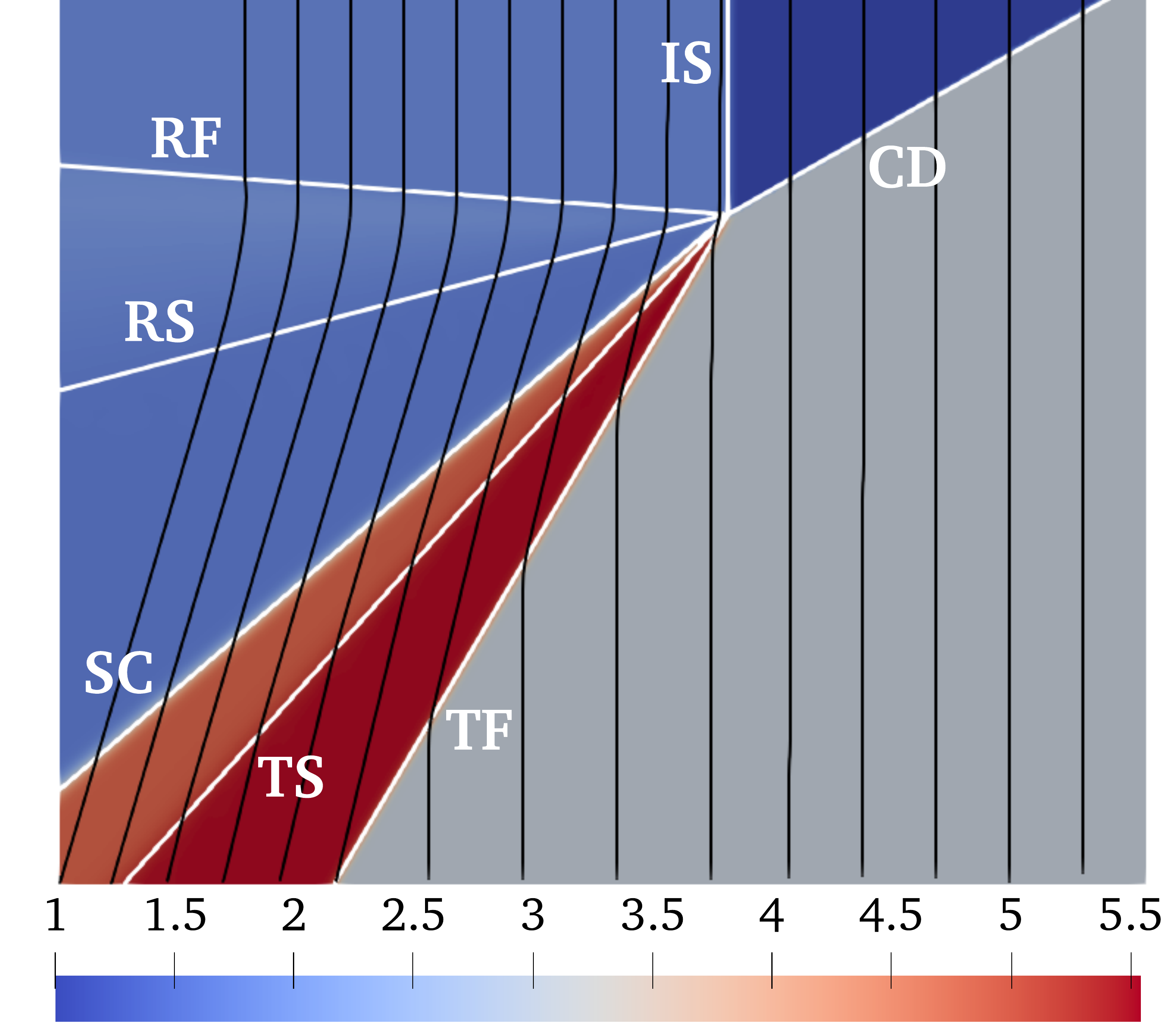}\label{fig:3.311a}
} 
\subfigure[]{ 
\includegraphics[scale=0.2]{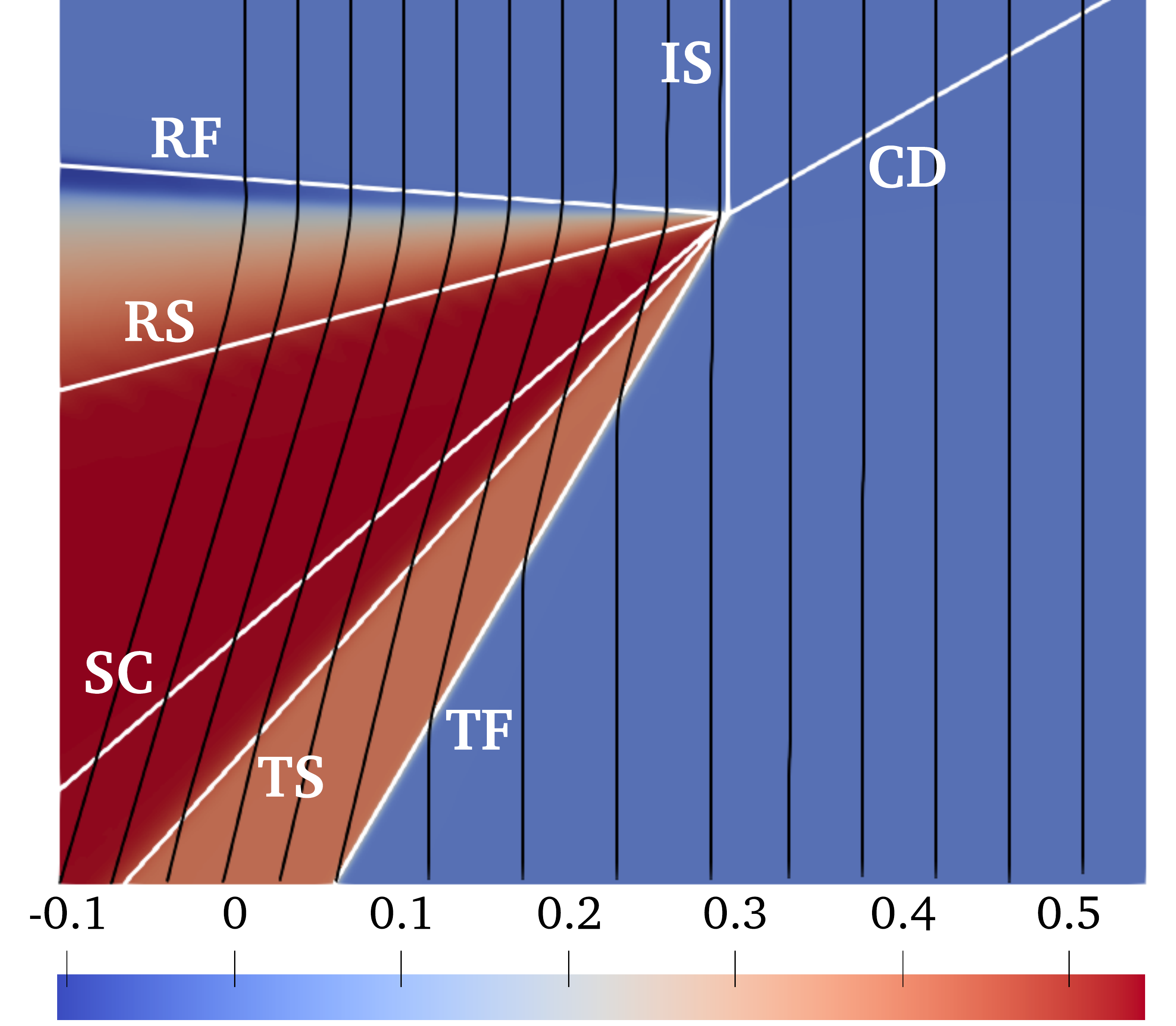}\label{fig:3.311b}
} 
\subfigure[]{ 
\includegraphics[scale=0.2]{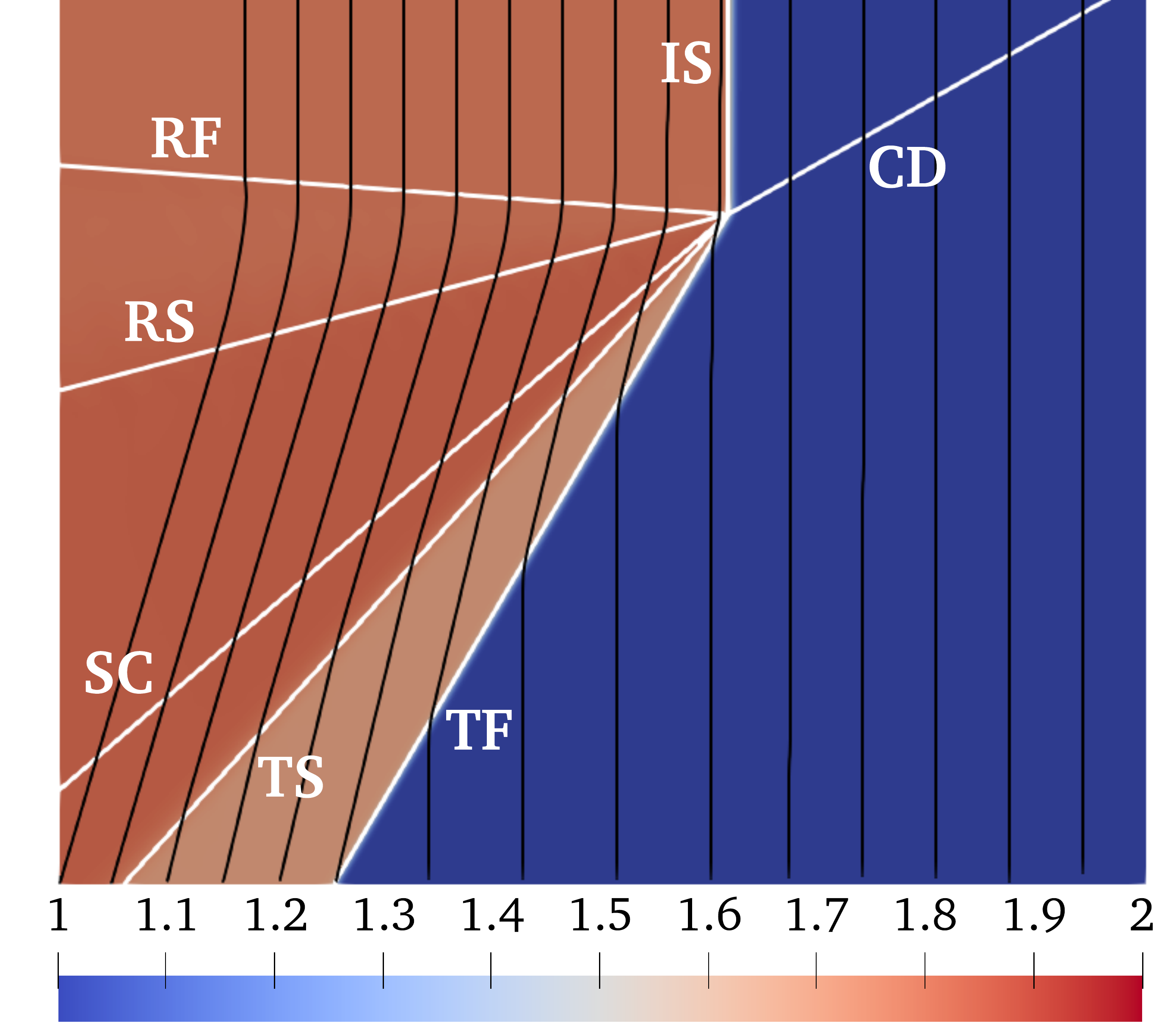}\label{fig:3.311c} 
} 
\caption{Analytical wave pattern of case \boldsymbol{$R_2$} overlaid on the numerical results displayed by density (a), $B_x$ (b) and $B_y$ (c) fields, respectively. Magnetic field lines are shown by black solid lines. }
\label{fig:3.311}       
\end{figure}

For a given set of parameters, the MHD shock refraction will transform regular into irregular refraction as the inclination interface angle $\alpha$ is decreased. 
First, we present the detailed wave structure of the case characterized with $M= 2, \beta =2, \gamma = 1.4, \eta = 3$ and $\alpha = \pi/6$, denoted as reference case \boldsymbol{$R_2$}. 
For this case, we follow the same analysis process for the case \boldsymbol{$R_1$} to show that $RS$ becomes a slow compound wave, $TS$ is a slow-mode expansion fan, $IS$, $RF$ and $TF$ are fast shocks. 
The compound wave $RS$ relevant here consists of a $2 \to 3=4$ intermediate shock, for which $v_{n2} = c_{SL2}$ , followed immediately downstream by a slow-mode expansion fan, where $c_{SL2}$ denotes the slow magnetosonic speed.  
Note that the case in the presence of parallel magnetic field and the hydrodynamic case, the refraction is irregular with $\alpha = \pi/6$, whereas the present case of the perpendicular magnetic field the refraction pattern is regular. 
Therefore, we are able to also obtain the analytical result in addition to numerical simulation results: both sets of results are plotted in Fig.~\ref{fig:3.311} with the white lines depicting the location of the waves from the analytical solution. 
The density field clearly displays the location of $TF$, $TS$ and $SC$ because of the strong density jump over these waves. The first two waves $TF$, $TS$  can also be seen in images of the x- and y-component of the magnetic field. 
On the other hand, the reflected waves $RF$ and $RS$ are weak, and we only discern these well in the image of the x-component of the magnetic field in Fig.~\ref{fig:3.311b}. 
The flow is compressed and the sign of $B_x$ is changed passing through the leading wavefront, followed immediately a slow-mode expansion fan which changes only the magnitude (not sign) of magnetic field and leads to state 3 at the trailing edge of the expansion fan which is part of the slow compound wave. 
The magnetic field lines are also overlaid in Fig.~\ref{fig:3.311} to show how the various shocks in the system deflect the field. 

\begin{figure}[htbp]   
\centering
\subfigure[]{
\includegraphics[scale=0.44]{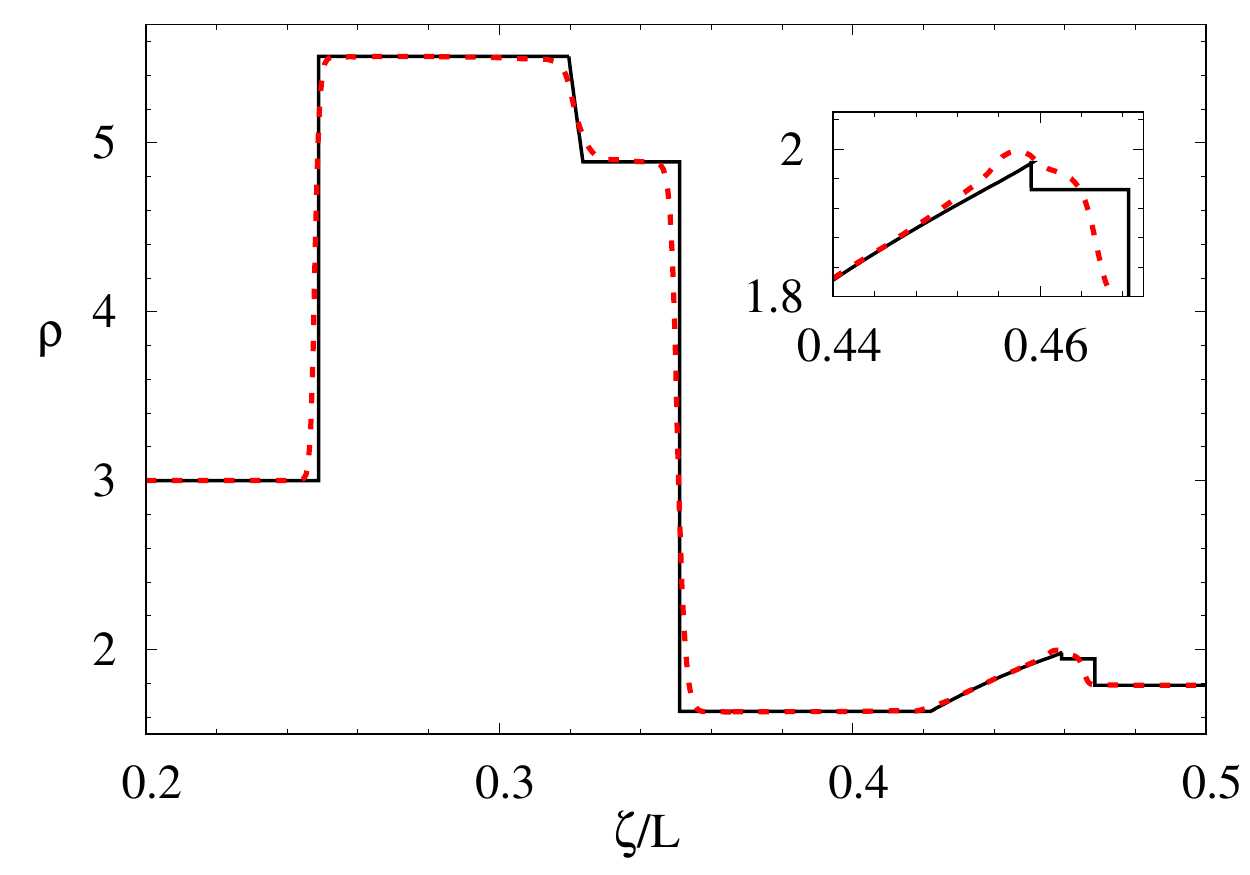}\label{fig:3.31a}
} 
\subfigure[]{ 
\includegraphics[scale=0.44]{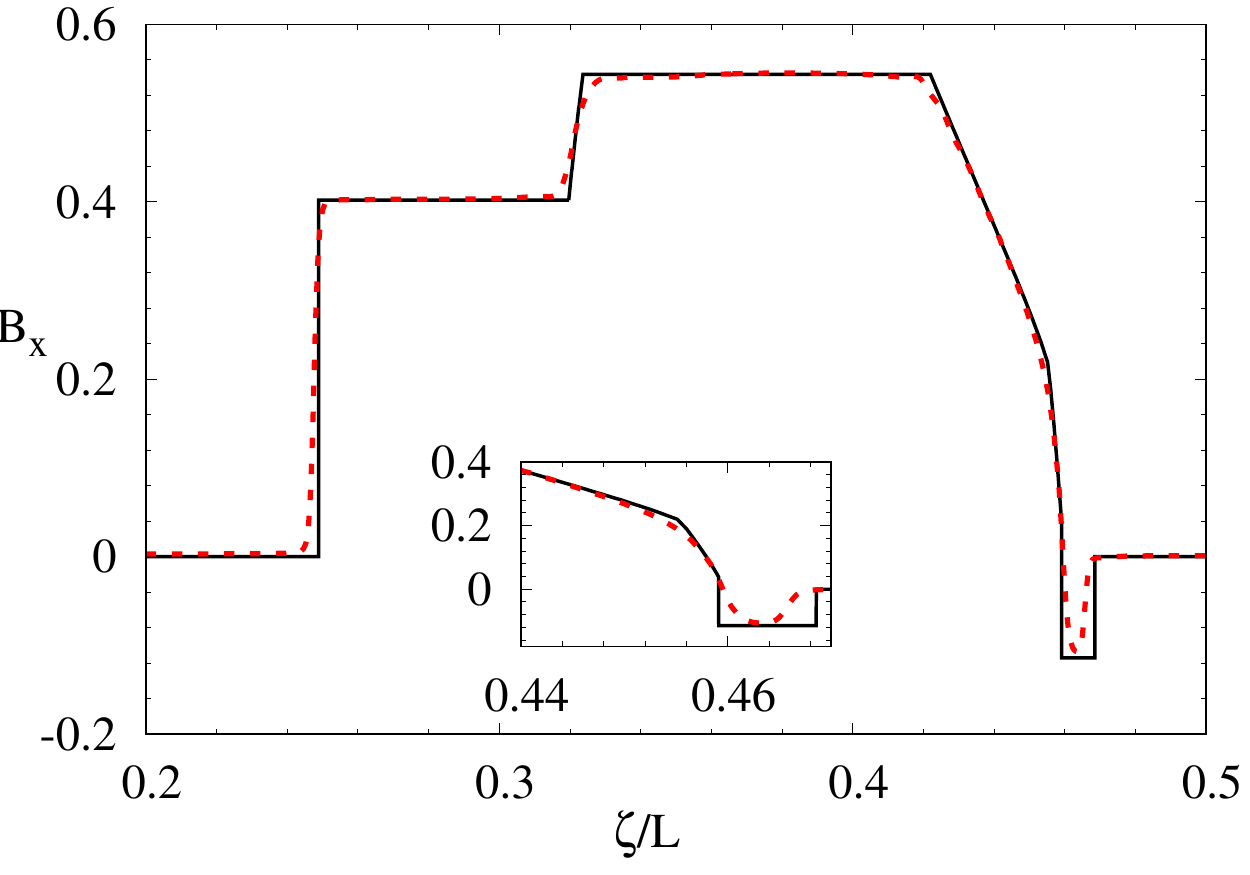}\label{fig:3.31b}
} 
\subfigure[]{ 
\includegraphics[scale=0.44]{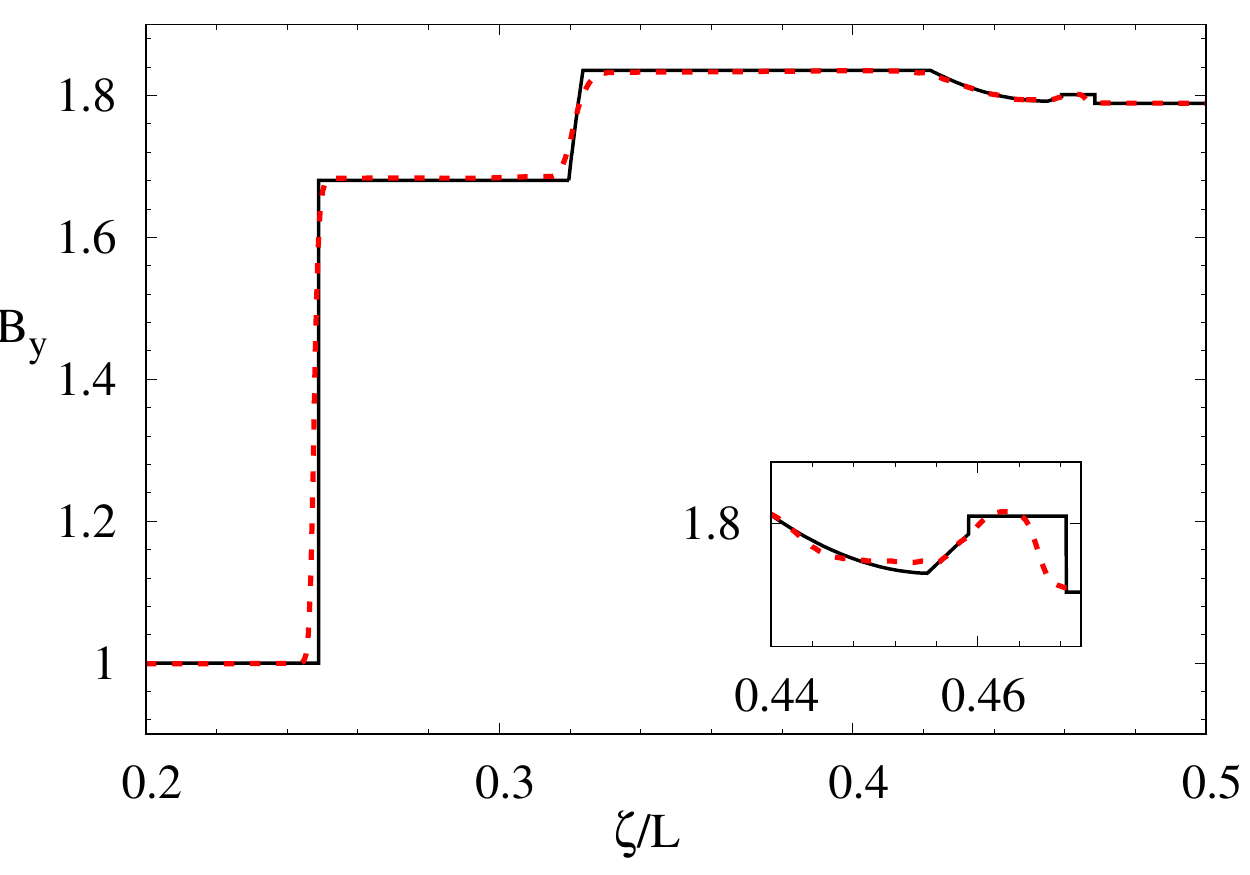}\label{fig:3.31c}
} 
\caption{Density $\rho$ (a) and magnetic field $B_x$ (b) $B_y$ (c) profiles from the numerical results at $\xi /L = 0.6667$ compared to profiles from analytical solution. Solid (dashed) line denotes analytical (numerical) solution.}
\label{fig:3.31}       
\end{figure}

In Fig.~\ref{fig:3.31}, we plot $\rho, B_x$ and $B_y$ profiles along a vertical line at $\xi /L = 0.6667$. 
From left to right in Fig.~\ref{fig:3.31a}, the discontinuities are as follows: fast shock $TF$, slow-mode expansion fan $TS$, SC, slow compound wave $RS$ and fast shock $RF$, respectively. 
In the analytical solution, we note the fine scale features of the slow compound wave  $RS$, the leading wavelfront of which is at $\xi/L = 0.459$ for $ \zeta/L = 0.667$. There appears to be good agreement between the analytical solution and the numerical results. 

\begin{figure}[htbp]   
\centering
\subfigure[]{
\includegraphics[scale=0.6]{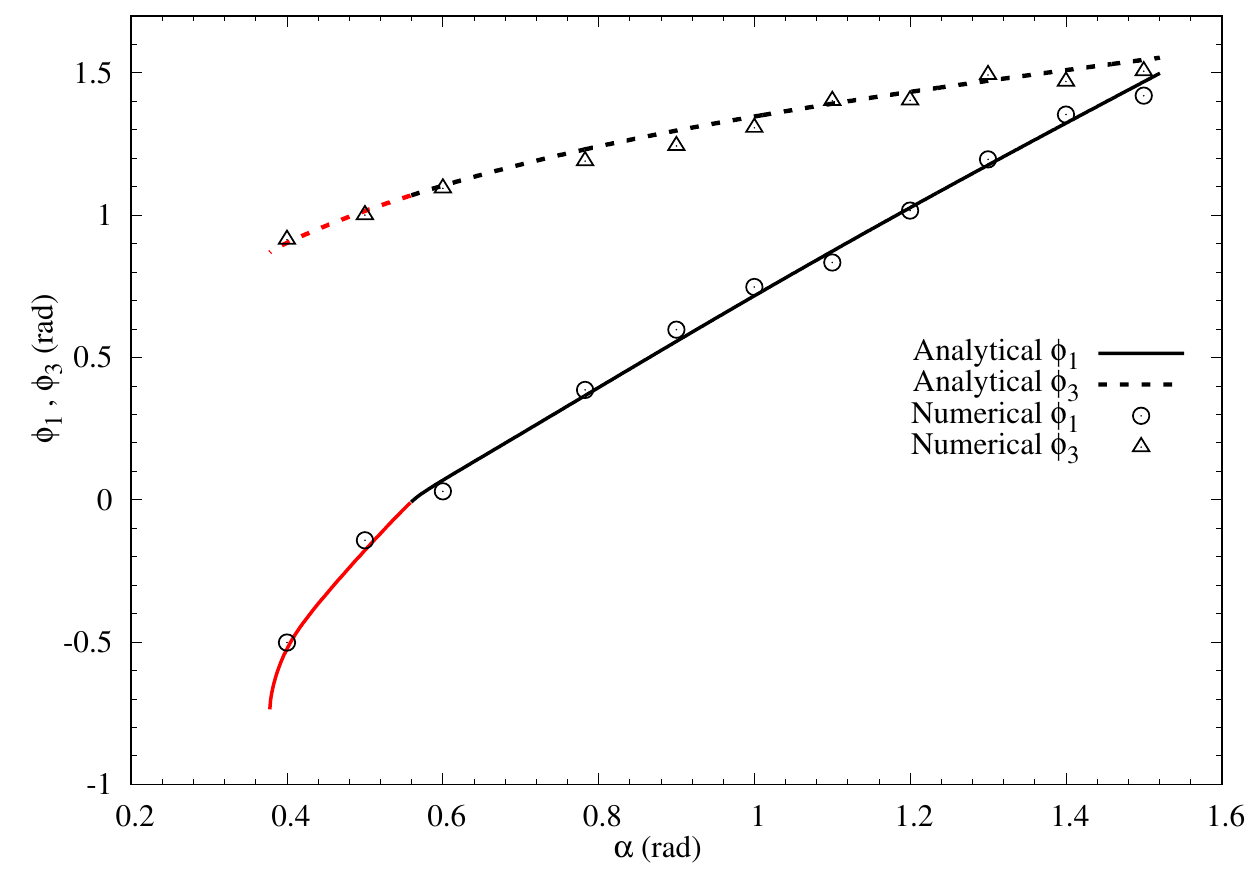}\label{fig:3.32a}
} 
\subfigure[]{ 
\includegraphics[scale=0.6]{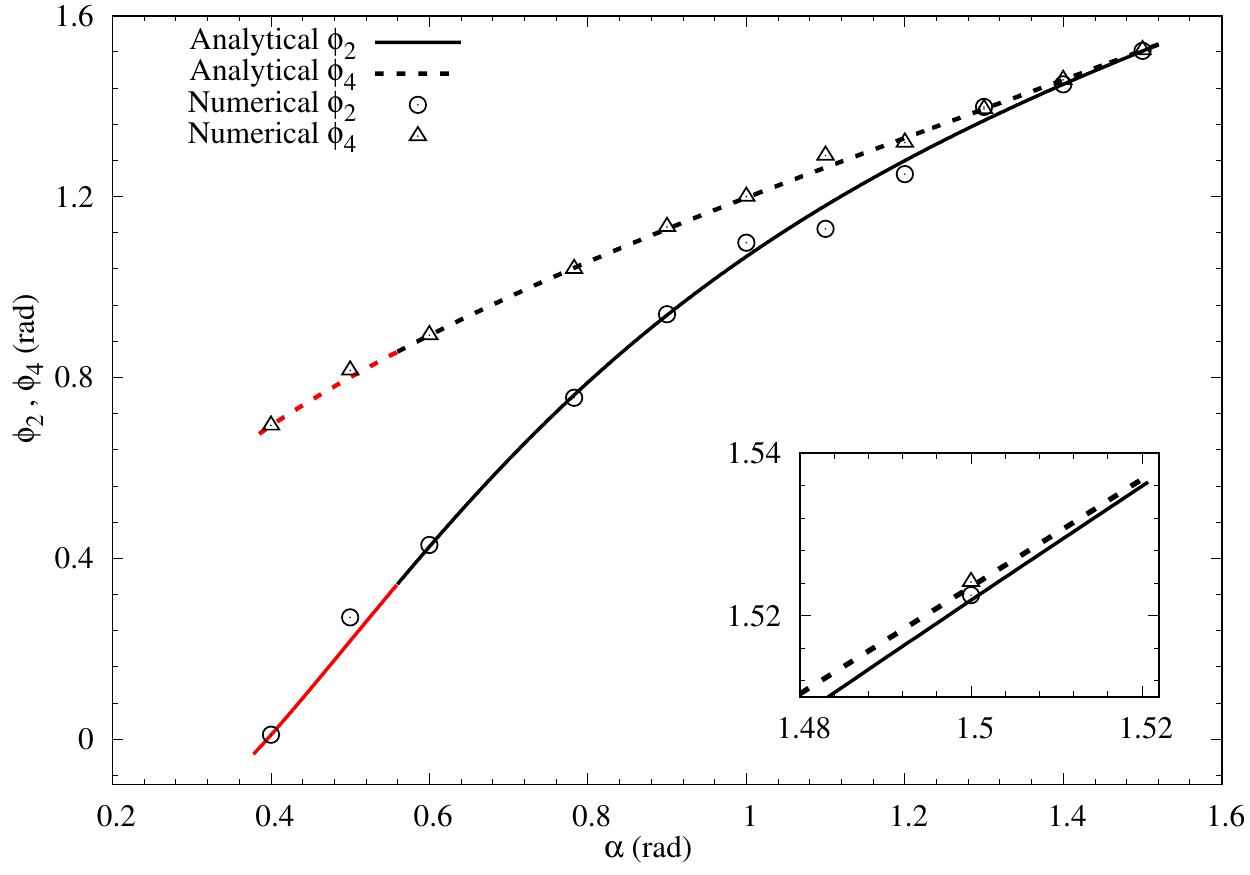}\label{fig:3.32b}
} 
\caption{(a) Fast shock angles $\phi_1$ and $\phi_3$ for $RF$ and $TF$, respectively; (b) slow-mode expansion fans or compound wave angles $\phi_2$ and $\phi_4$ for $RS$ and $TS$, respectively. The black (red) curve denote that the RS is slow-mode expansion fans (slow-mode compound wave).}
\label{fig:3.32}       
\end{figure}

We now investigate the wave pattern by varying $\alpha \in (0.378, 1.521)$ and fixing $M= 2, \beta =2, \gamma = 1.4$ and $\eta = 3$. In Fig.~\ref{fig:3.32a}, we show that the location angles $\phi_1$ of $RF$ and $\phi_3$ of $TF$ increase as $\alpha$ is increased. The transition of $RS$ occurs at $\alpha = 0.559$, the black (resp. red) curve in the figure corresponds to $RS$ being a slow-mode expansion fan (resp. slow compound wave). 
The angle $\phi_3$ diminishes smoothly as $\alpha$ decreased over the range of $\alpha$ that was examined. 
For $RF$, $\phi_1$ decreases linearly as $\alpha$ decreased when the $RS$ is slow-mode expansion fan, whereas it decreases more strongly and the $RF$ location changes the sign ($\phi_1 < 0$) when $RS$ transitions  to a slow-mode compound wave ($\alpha < 0.559$). We further note that $RF$ location angle increases strongly towards the positive y-axis direction up to $\phi_1 =-0.736439$ (negative meaning the angle is now measured clockwise from the negative x-axis) at $\alpha = 0.378$. 
In Fig.~\ref{fig:3.32b}, the location angle $\phi_2$ of $RS$ and $\phi_4$ of $TS$ diminish smoothly as $\alpha$ is decreased, and there is no obvious change of angle evolution in the transition region. For the maximum value $\alpha = 1.521$ in this study, $\phi_2$ tends to be close to $\phi_4$ as shown in the zoomed in region in Fig.~\ref{fig:3.32b}. 

\begin{figure}[htbp]   
\centering
\subfigure[]{
\includegraphics[scale=0.6]{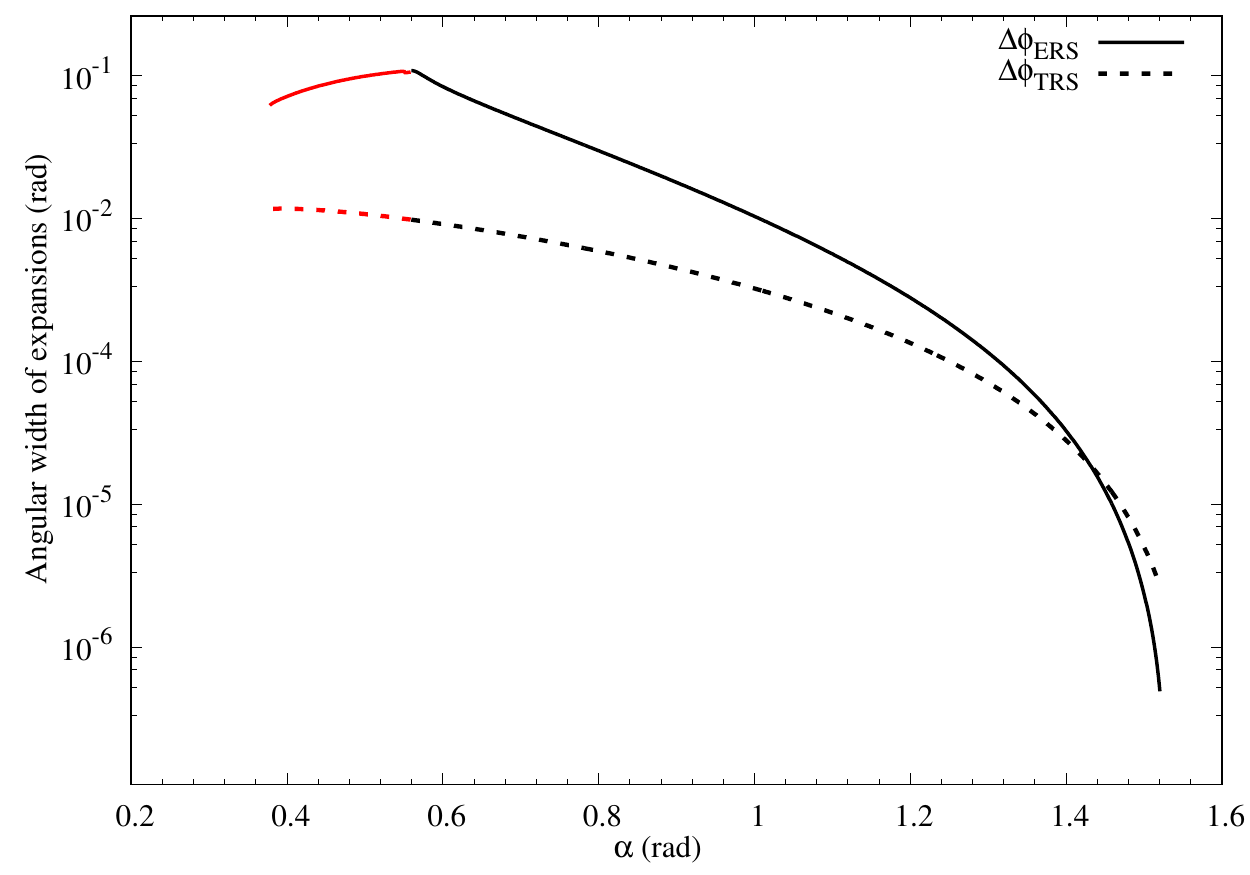}\label{fig:3.33a}
} 
\subfigure[]{ 
\includegraphics[scale=0.6]{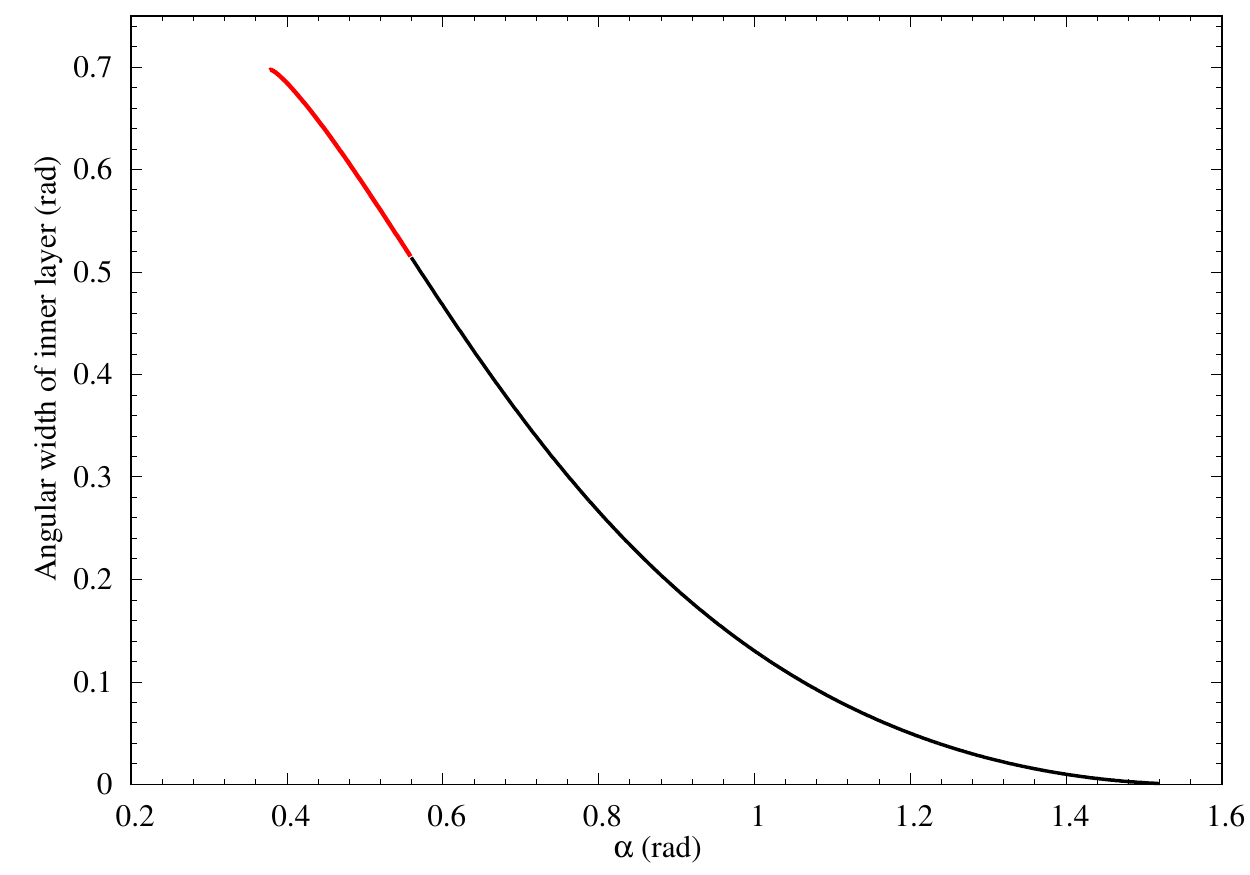}\label{fig:3.33b}
} 
\caption{(a) Angular width of slow-mode expansion fans, $\Delta \phi_{ERS}$ and $\Delta \phi_{ETS}$ for $RS$ and $TS$, respectively; (b) angular width of inner layer. The black (red) lines stand that the RS is slow-mode expansion fans (slow-mode compound wave).}
\label{fig:3.33}       
\end{figure}

Fig.~\ref{fig:3.33a} shows that as $\alpha$ decreased, the angular width of slow-mode expansion fan $\Delta \phi_{ETS}$ increases smoothly. 
For the reflected $RS$ wave, the angular width $\Delta \phi_{ERS}$ increases as $\alpha$ decreases if $RS$ is a slow-mode expansion fan. 
On the other hand, $\Delta \phi_{ERS}$ decreases as $\alpha$ decreases when $RS$ becomes a slow compound wave. 
The angular width of inner layer increases smoothly with decreasing of $\alpha$ as shown in Fig.~\ref{fig:3.33b}. 
For $\alpha < 0.378$, the analytical solution does not exist for the chosen parameters ($M= 2, \beta =2, \gamma = 1.4$ and $\eta = 3$). This is because shock refraction transitions to an irregular pattern. We now explore the critical angle $\alpha_{crit}$ of regular-irregular transition by varying the magnetic strength.  

\begin{figure}[htbp]   
\centering
\subfigure[]{
\includegraphics[scale=0.6]{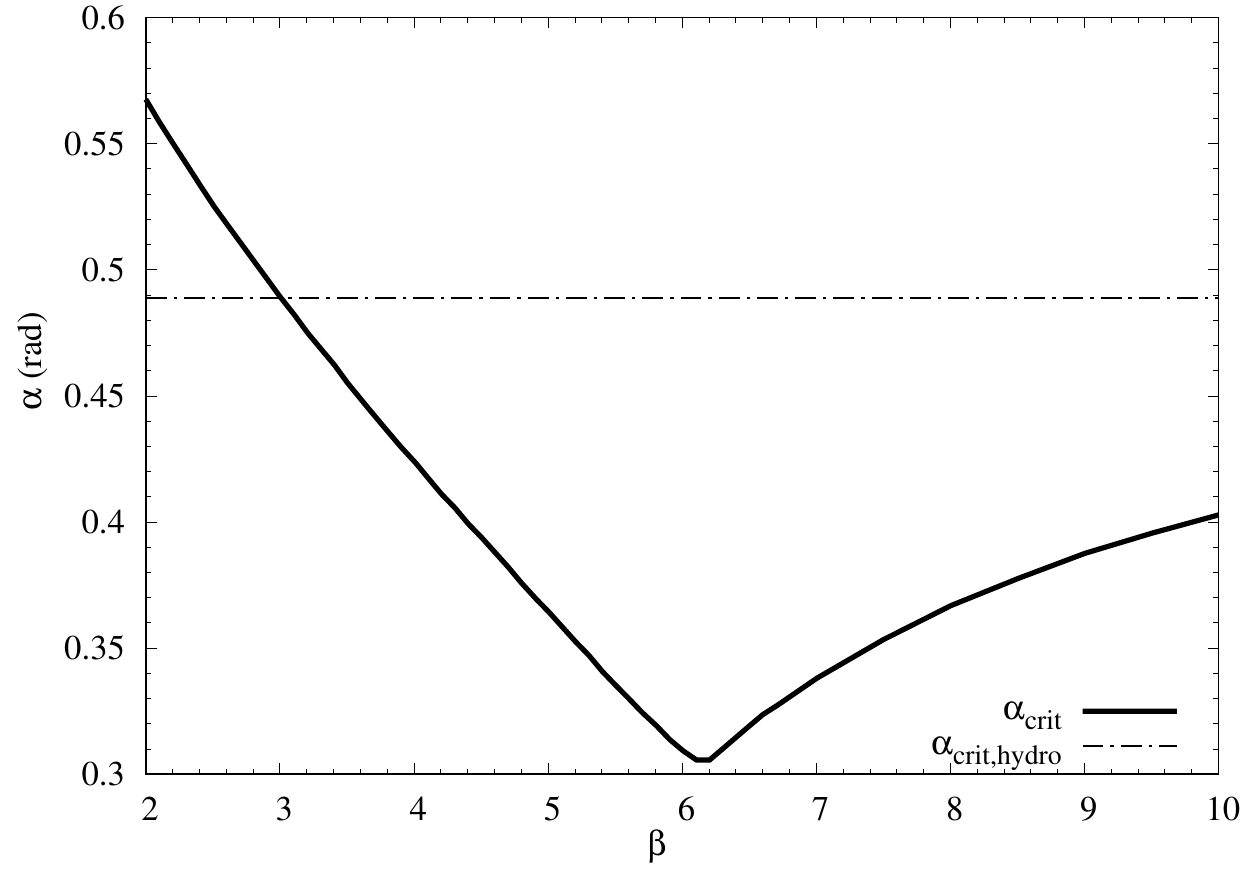}\label{fig:3.34a}
} 
\subfigure[]{ 
\includegraphics[scale=0.6]{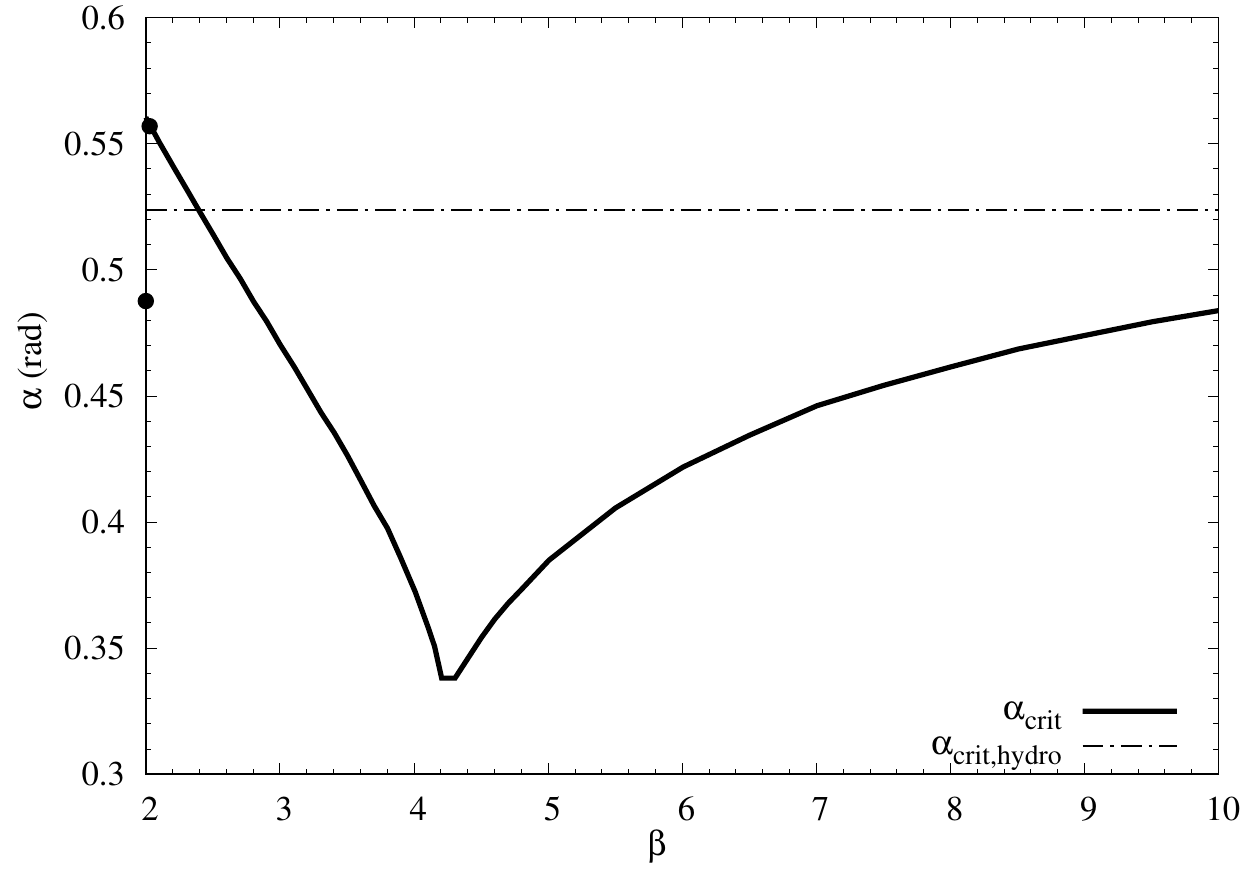}\label{fig:3.34b}
} 
\subfigure[]{ 
\includegraphics[scale=0.6]{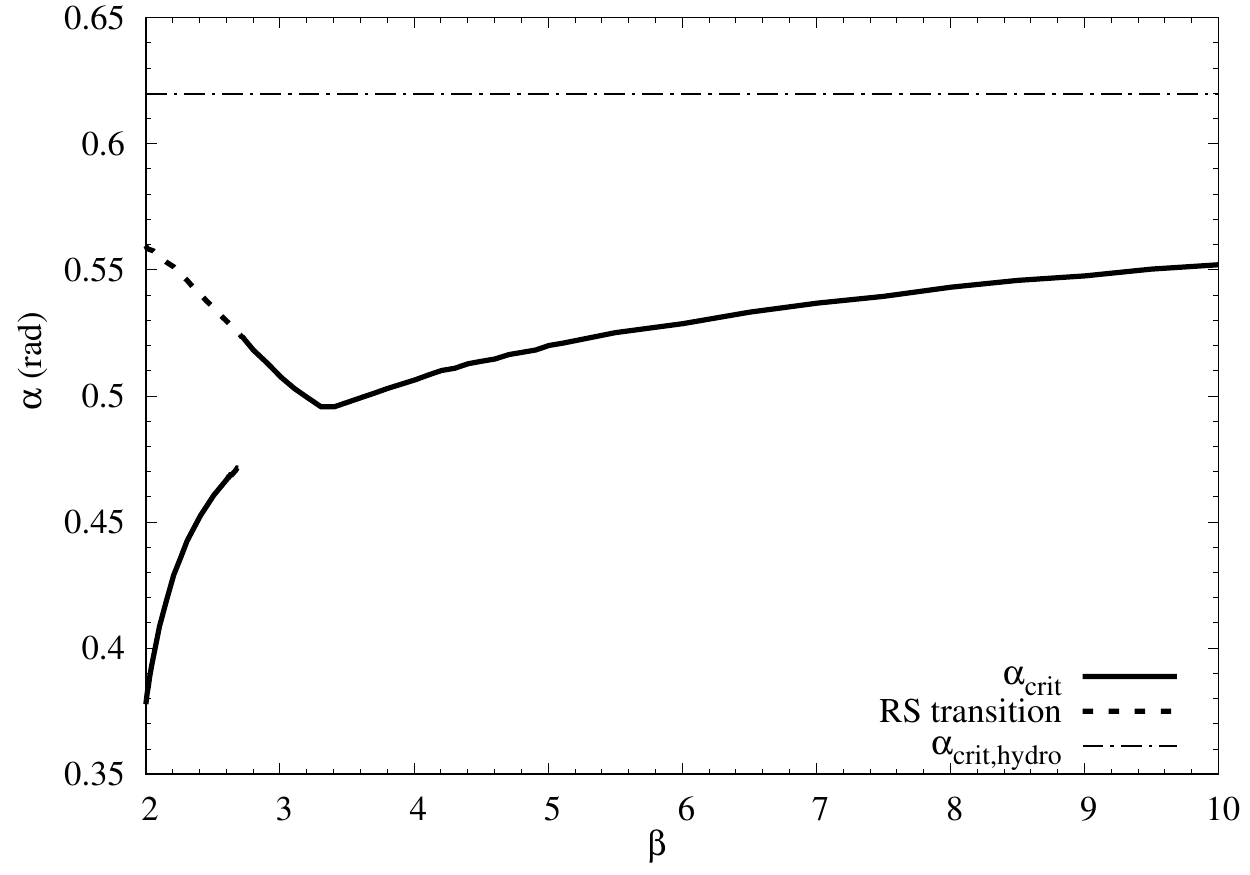}\label{fig:3.34c}
} 
\subfigure[]{ 
\includegraphics[scale=0.6]{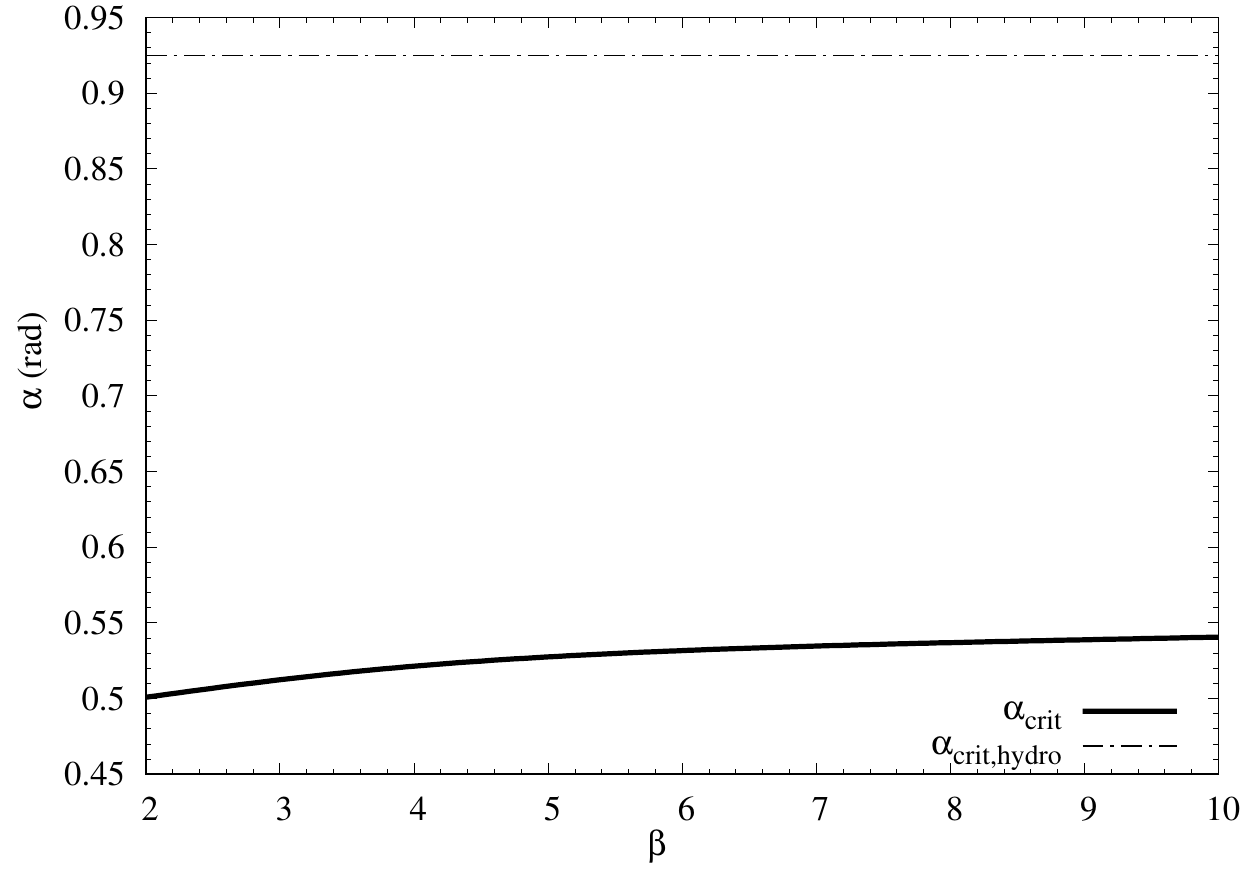}\label{fig:3.34d}
} 
\caption{Critical angle $\alpha_{crit}$ of regular to irregular transition for the MHD shock refraction problem. (a) $M=1.35$, (b) $M=1.5$, (c) $M=2$, (d) $M=5$.}
\label{fig:3.34}       
\end{figure}

For the chosen parameter set ($M = 2, \eta = 3, \gamma = 1.4$), the critical angle at which transition from regular to irregular refraction takes place is $\alpha_{crit}$ is around $35.5^{\circ}$ for hydrodynamic shock refraction $(\beta^{-1} = 0)$.
In the studied range, we find that the presence of an initially perpendicular magnetic field delays the transition to irregular refraction compared to the hydrodynamic case, see in Fig.~\ref{fig:3.34c}. 
In the presence of a strong magnetic field, $\alpha_{crit}$ increases strongly as magnetic field weakens until $\beta = 2.7$. 
It is found that the transition angle of $RS$ from slow-mode expansion fan to compound wave decreases as $\beta$ is increased, and the reflected wave $RS$ is slow compound wave just before the wave pattern transitions to an irregular pattern in this range.
In range of $\beta \in (2.7, 3.406)$, $\alpha_{crit}$ suddenly increases and is identical to the RS transition angle which in turn also deceases as $\beta$ is decreased.
This range is somewhat unique where the $\alpha_{crit}$ versus $\beta$ plot is discontinuous, an undershoot of $\alpha_{crit}$ is found at $\beta = 3.406$. 
Beyond the above ranges, the RS transition angle cannot be analytically calculated, since it is less than $\alpha_{crit}$ so that RS is a slow-mode expansion fan when the wave pattern becomes irregular. To explore this somewhat anomalous behavior of critical angle $\alpha_{crit}$ versus $\beta$, we also compute the critical angle for three other Mach numbers, $M=1.35, \ 1.5$ and 5, for which the critical angle is plotted in Fig.~\ref{fig:3.34a}, Fig.~\ref{fig:3.34b} and Fig.~\ref{fig:3.34d}, respectively. 
In the critical angle $\alpha_{crit}$ versus $\beta$ evolution, the discontinuous region is expanded and the undershoot occurs at larger $\beta$ as Mach number decreases (see $M=1.35, \ 1.5$  cases). The undershoot of $\alpha_{crit}$ occurs at $\beta = 4.207$ and $6.106$ for $M= 1.5$ and $1.35$, respectively. The $\alpha_{crit}$ is $0.4876$ at $\beta=2$ and suddenly increases to $0.5569$ at $\beta = 2.029$ for $M=1.5$ (points in Fig.~\ref{fig:3.34b}), while this feature is not observed in the studied range for $M=1.35$. 
With an increase in Mach number, $\alpha_{crit}$ versus $\beta$ plot is smooth as for the $M=5$ case. 
It is evident that the transition from regular to irregular refraction is a complex phenomenon and no clear trend, especially at weaker Mach numbers, is apparent. A more thorough exploration of the parameter space is left for the future. 
%
%


\section{\label{sec:4} Results: Irregular Refraction}
\begin{figure}[h]   
\centering
\subfigure[]{
\includegraphics[scale=0.26]{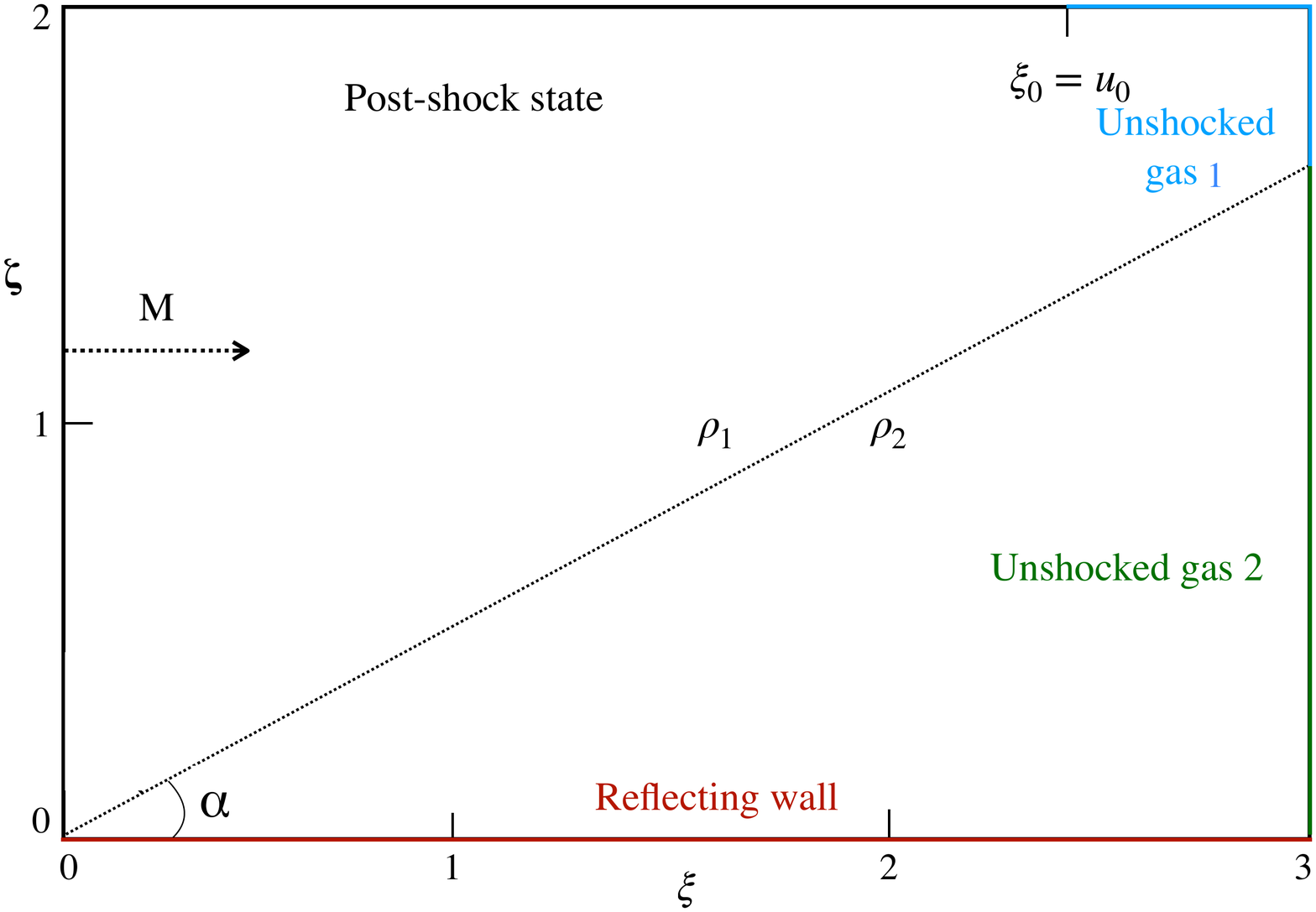}\label{fig:3.4a}
} 
\subfigure[]{ 
\includegraphics[scale=0.26]{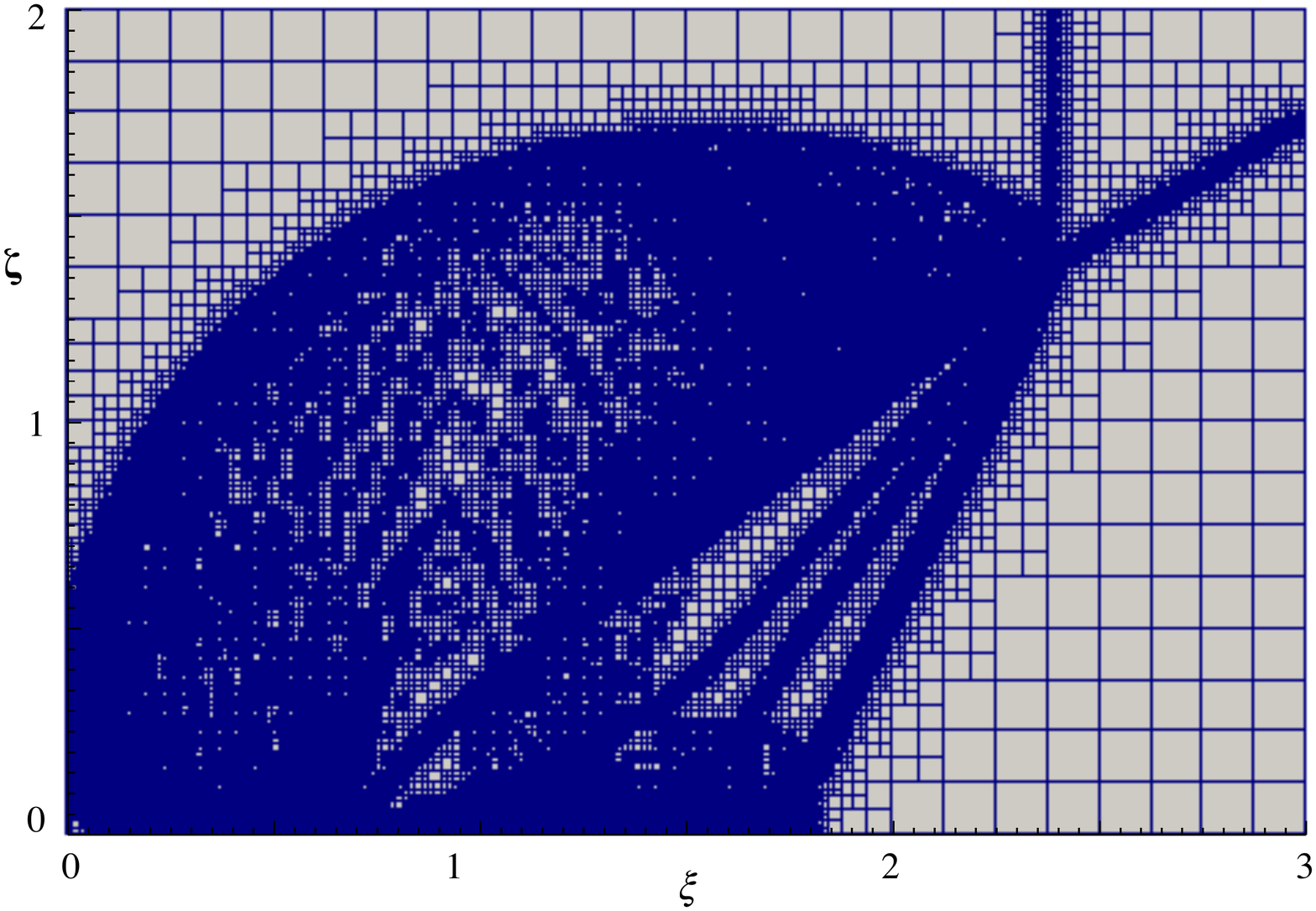}\label{fig:3.4b}
} 
\caption{(a) Boundary conditions and initial guess for the two-dimensional shock contact-discontinuity interaction; (b) AMR Mesh structure with 9 levels of refinement for the shock refraction with the presence of a parallel magnetic field at convergence.}
\label{fig:3.4}       
\end{figure}

Presently, we solve the self-similar formulation of ideal GLM-MHD equations (Eq.~\ref{eq:2.23}) to investigate the irregular MHD shock refraction, including both the perpendicular and parallel orientations of the initially applied magnetic fields, as well as the hydrodynamic case.  
We compute the specific cases with parameters $M=2, \eta = 3, \gamma = 1.4$, and vary $\beta$ and $\alpha$. 
For the self- similar solution the boundary conditions and the initial guess are depicted in Fig.~\ref{fig:3.4a}. 
The pressure on either side of the contact-discontinuity is unity. The state behind the incident shock is obtained by the Rankine-Hugoniot jump condition, and the shock is moving with the speed of $u_0=M~c_f$, where $c_f$ is the fast magnetosonic sound speed. 
The computational domain (now in velocity coordinates after the self similar transformation) $\Omega = [0,3]\times [0,2]$ is discretized with mesh refinement utilizing 9 levels (base mesh cell size is unity ), yielding an effective uniform mesh resolution of $1536 \times 1024$. An example of AMR mesh structure is shown in Fig.~\ref{fig:3.4b}.

\begin{figure}[] 
\includegraphics[scale=0.25]{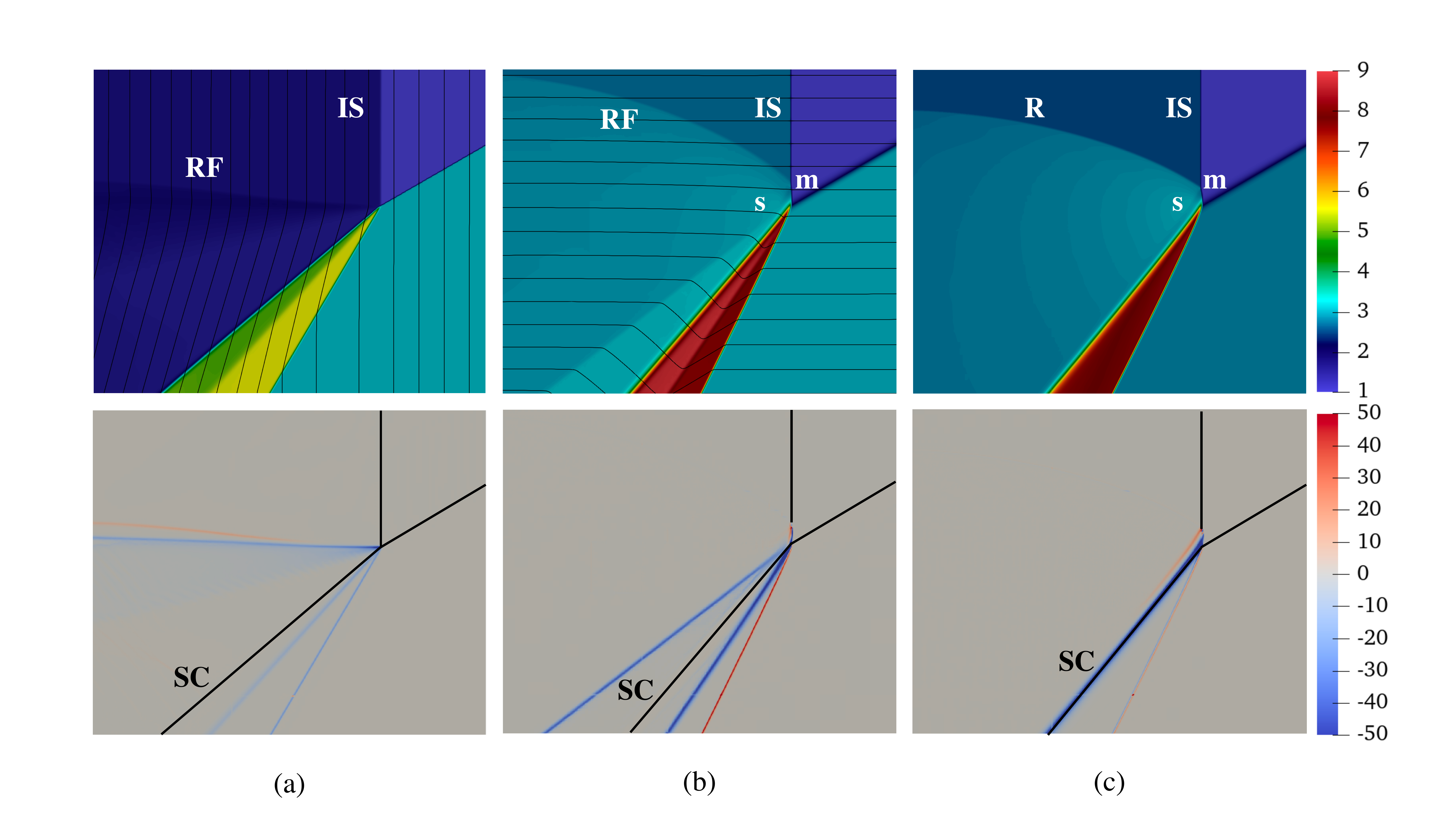}   
\caption{\label{fig:3.41}Density (first row) and vorticity (second row) fields of the shock refraction with $\alpha = \pi/6$. (a) case with the presence of perpendicular magnetic field $\beta = 2$, (b) case with the presence of parallel magnetic field $\beta = 2$, (c) hydrodynamic case $\beta^{-1} = 0$.
}   
\end{figure}

\begin{figure}[] 
\includegraphics[scale=0.25]{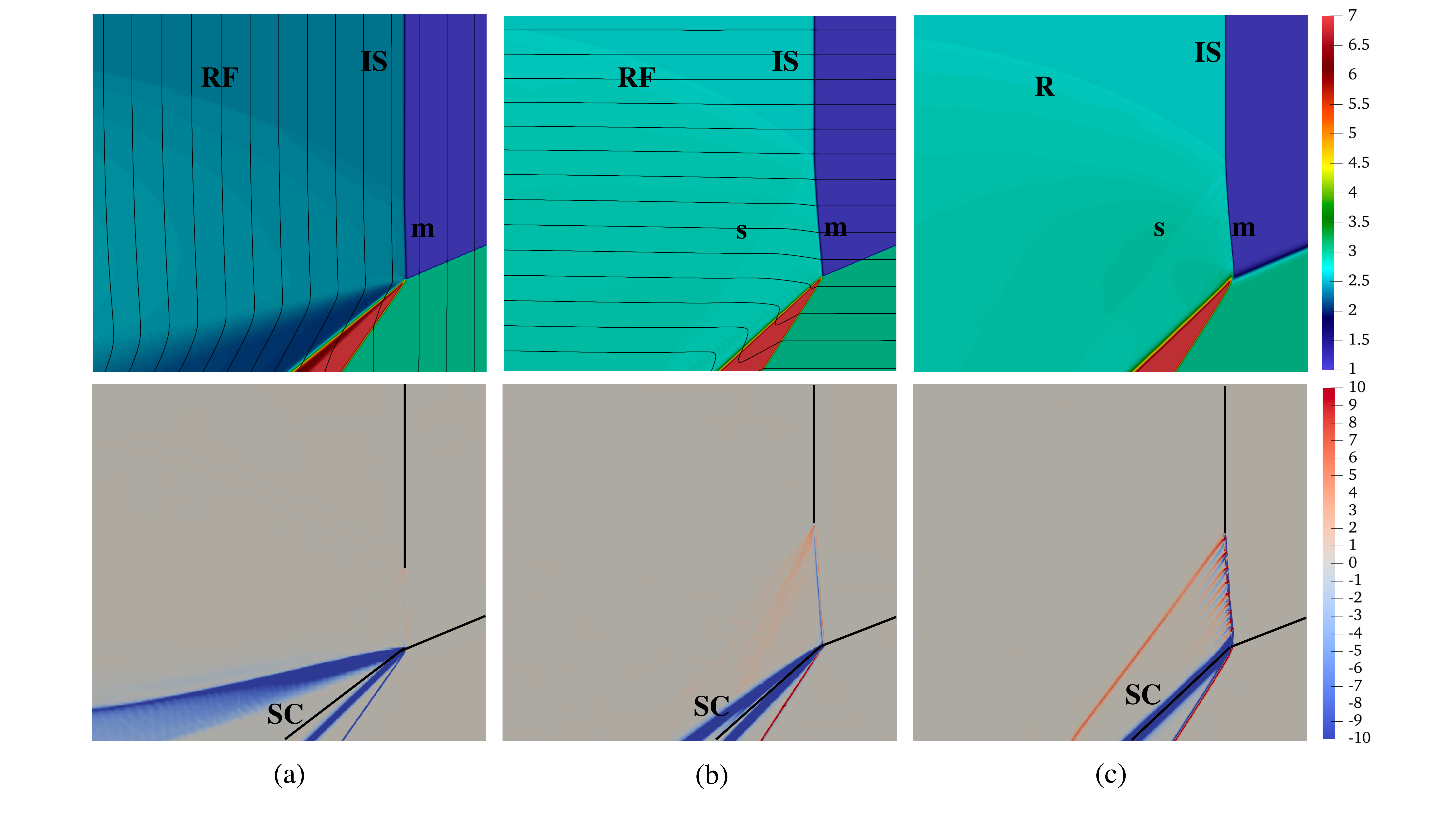}   
\caption{\label{fig:3.42}Density (first row) and vorticity (second row) fields of the shock refraction with $\alpha = 0.4$. (a) case with the presence of perpendicular magnetic field $\beta = 8$, (b) case with the presence of parallel magnetic field $\beta = 8$, (c) hydrodynamic case $\beta^{-1} = 0$.
}  
\end{figure}

The self-similar solutions with $\alpha = \pi/6$ are shown in Fig.~\ref{fig:3.41}, in which the left/right sequence is the case of the perpendicular, parallel magnetic field $(\beta = 2)$ and hydrodynamic case $(\beta^{-1} = 0)$, respectively. 
For the hydrodynamic case, the Mach stem $m$ connects the reflected shock $R$ to the point where $m$ intersects the density interface.  
A second shear layer $s$ emerges from the triple-point where $IS$, $R$ and the Mach stem $m$ intersect. This system is called single Mach-reflexion irregular refraction $MRR$, which has been observed in the hydrodynamic experiments~\cite{Henderson1978}. 
The irregular shock structure produced in the parallel magnetic field case is  the MHD equivalent of the single Mach reflexion $MRR$. The reflected fast shock $RF$ no longer intersects the density interface, and is connected to the density interface, transmitted shocks and $RS$ by the MHD equivalent of a Mach stem $m$. 
The changes in flow properties across the MHD Mach stem $m$ are consistent with it being a fast mode MHD shock. Fig.~\ref{fig:3.42} (b) shows that it compresses the flow without changing sign of the magnetic field. 
Furthermore, the vorticity generated on the vicinity of shocked contact $SC$ is transported onto the MHD waves, as in the case of regular refraction. 
This transport of baroclinic vorticity from the interface prevents the shear instability that causes the roll-up of the interface in the hydrodynamic case. These results indicate that the mechanism by which a magnetic field suppresses the MHD RMI is valid independent of whether regular or irregular refraction occurs at the density interface (for a discussion of this we refer to the work of Samtaney~\cite{Samtaney2003} and Wheatley \etal~\cite{Wheatley2005JFM}).
Moreover, the second shear layer $s$ is not present and instead the fast mode Mach stem $m$ is a vortex sheet. 
On the other hand, the resulting shock structure of the case in the presence of perpendicular magnetic field is regular, the analytical and numerical solution is shown in detail in Fig.~\ref{fig:3.311} and Fig.~\ref{fig:3.31}. 
The vorticity on the vicinity of shocked contact $SC$ is transported onto the MHD waves (slow compound wave $RS$ and slow expansion fan $TS$), as in the parallel magnetic field case. 
For this specific set of parameters, it is clearly shown that the presence of perpendicular magnetic field delays the regular-irregular transition, which is consistent with the analytical conclusion presented in Section~\ref{subsec:3.3}, whereas the presence of parallel magnetic field suppresses the second shear layer and leads to the presence of a Mach stem with shear.  

We now compute the above three cases with the inclination angle $\alpha = 0.4$ and $\beta = 8$. 
The parameter combination of $\beta = 8$ and $\alpha = 0.4$ leads the refraction pattern to be irregular for all three cases. 
Fig.~\ref{fig:3.42} shows the irregular refraction wave pattern with this specific choice of parameters. 
For the hydrodynamic case $(\beta^{-1} = 0)$, the incident and Mach-stem shocks $IS$ and $m$ appear as a continuous wave which is convex-forwards along the segment formerly occupied by the incident shock $IS$. 
The reflected shock $R$ disperses into a band of wavelets. 
In addition to the usual second shear layer $s$ stemming from the triple point where $IS$, $R$ and the Mach stem $m$ intersect, there is another band of shear layers visible in the hydrodynamic case. 
This system is called as convex-forwards irregular refraction $CFR$ by Henderson \etal ~\cite{Henderson1978}. 
In the MHD cases, the band of shear layers that emerges from Mach stem completely vanishes for both orientations (parallel and perpendicular) of the magnetic field. 
Here $s$ can be considered as a fast mode MHD shock, since the flow across this wave is compressed without changing sign of magnetic field.  
Hence with the specific parameters $\alpha = 0.4$ and $ \beta = 8$, the irregular shock structures produced in the presence of perpendicular or parallel magnetic fields cases appear also to be the MHD equivalent of the single Mach reflexion refraction (MRR), while it is $CFR$ type for the hydrodynamic case.
The presence of a magnetic field weakens the Mach stem compared to the hydrodynamic case: the perpendicular magnetic field attenuates these features more effectively than the parallel field case. 
The Mach stem in the perpendicular magnetic field case lags behind the other two cases. 
Furthermore, the vorticity deposited on the vicinity of $SC$ is also transported on the MHD waves for irregular refraction in the perpendicular magnetic field case. This has implications on the suppression of the RMI, in the sense that even for irregular refraction the vorticity is not on the density interface and one expects the instability suppression to still take place. 

\begin{figure}[] 
\includegraphics[scale=0.25]{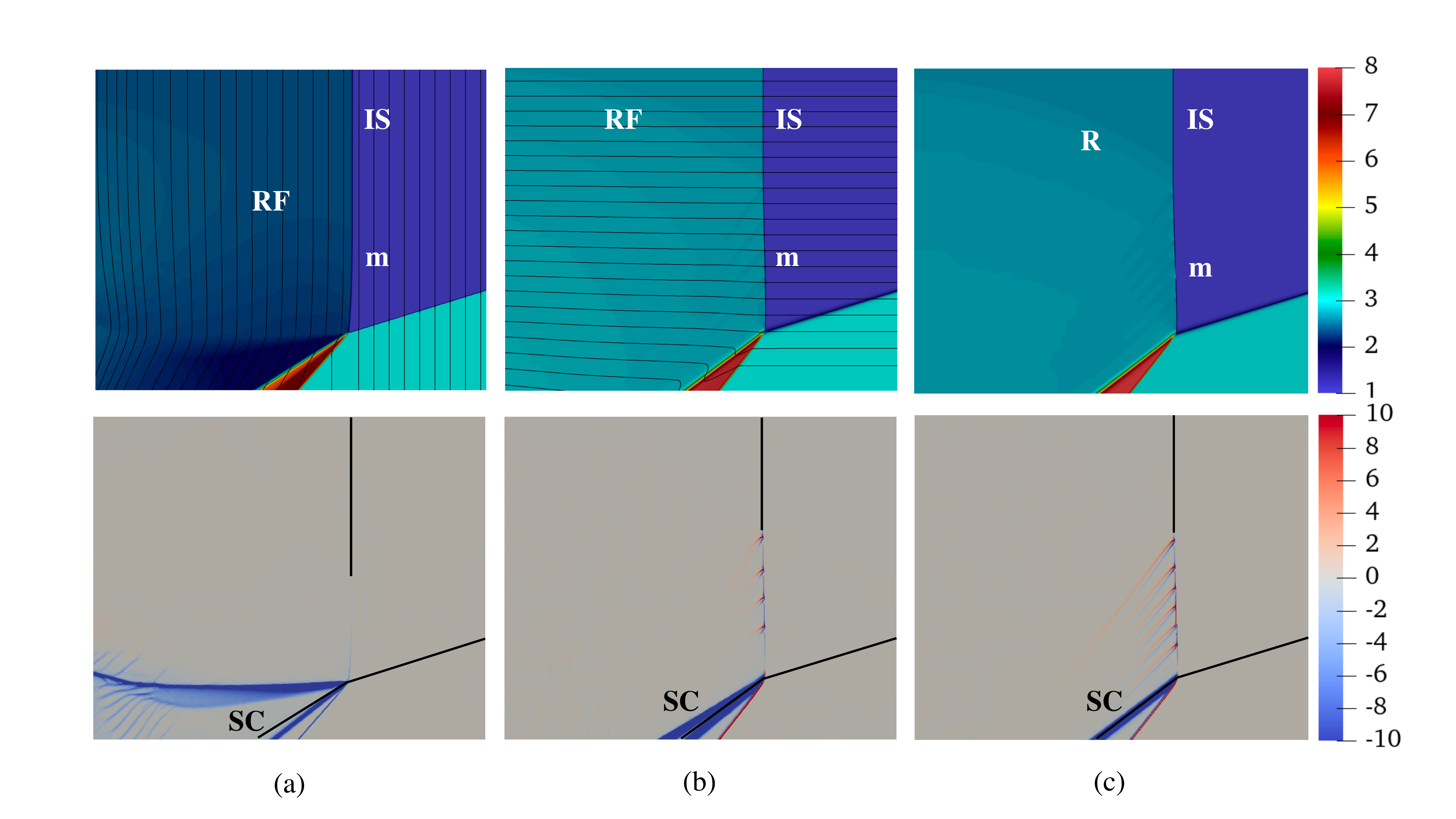}    
\caption{\label{fig:3.43}Density (first row) and vorticity (second row) fields of the shock refraction with $\alpha = 0.3$. (a) case with the presence of perpendicular magnetic field $\beta = 8$, (b) case with the presence of parallel magnetic field $\beta = 8$, (c) hydrodynamic case $\beta^{-1} = 0$.
} 
\end{figure}

We further decrease the inclination interface angle to $\alpha = 0.3$. Now three irregular flow structures appear as shown in Fig.~\ref{fig:3.43}. 
For the hydrodynamic case, the reflected shock has apparently dispersed into a band of wavelets, the band of shear layers emerges from Mach stem $m$ which is convex-forwards, and the irregular wave structure is identified to the $CFR$ type (similar to the one for $\alpha = 0.4$). 
In the presence of magnetic fields, the reflected shock has apparently dispersed into a band of wavelets and the wave structure is considered as the MHD equivalent of $CFR$ type that occurs in hydrodynamic case.
These hydrodynamic $CFR$ wave structures have been observed in experiments, see Figure 11 and Figure 12 in~\cite{Henderson1978}.
In the MHD equivalent of $CFR$ irregular structure, the Mach stem is less convex-forwards than the corresponding case with $\alpha = 0.4$ ($MRR$ type). 
The changes in flow properties across the MHD Mach stem and waves named as band of shear layers in hydrodynamic case are also consistent with they being fast mode MHD shocks.  
As the conclusion for the cases with $\alpha = 0.4$, it is found that the presence of perpendicular magnetic field weakens more efficiently the Mach stem than the parallel magnetic field case.  

\section[*]{\label{sec:5} Conclusion}
In summary, the present work investigates the ideal MHD flow structure produced by the refraction of a shock at an oblique planar density interface assuming the flow to be strongly planar (i.e., no velocity or component of the magnetic field in the z-direction).  
We consider the cases in the presence of magnetic fields, which are initially perpendicular and parallel to the motion of incident shock.
We employ an iterative procedure to obtain the analytical solution to MHD regular shock refraction. The analysis is restricted to cases where the initially applied magnetic field is oriented perpendicular to the direction of shock wave propagation. For computing the flow fields numerically, we solve a boundary value problem stemming from a self-similar transformation of ideal GLM-MHD equations. The numerical simulations are used for both regular and irregular shock refractions. 
Specifically, we analytically investigate the regular solution to the cases with the presence of perpendicular magnetic fields, and discover that the wave structure consists of three fast shocks ($IS, RF$ and $TF$), a slow-mode expansion fan ($TS$), and a slow-mode expansion fan or slow compound wave for $RS$ by varying the magnetic field magnitude and the inclination interface angle. 
The inclination interface angle of $RS$ transition from slow-mode expansion fan to slow compound wave decreases as magnetic field magnitude decreases. 
In general, the analytical solutions agree well with the numerically computed ones. We plot the deviation of the wave angles from the corresponding hydrodynamic regular refraction case. The angular extent of the slow mode expansion fans reach a maximum for a particular value of the magnetic field ($\beta \approx 2.3$) and then gradually decrease with increase in $\beta$. The angular width of the inner layer, the angles of the fast mode shocks ($RF$ and $TF$), and the angular extents of the slow mode expansion fans ($RS$ and $TS$) scale with $\beta^{-1/2}$, i.e. linearly with the strength of the applied magnetic field for large $\beta$. 
We also quantified the critical angle where the transition from regular to irregular refraction takes place. 
At $M=2$, perpendicular magnetics suppress the transition compared to the corresponding hydrodynamic case, although the critical angle versus $\beta$ plot is not monotonous. 
This non-monotonic is expanded with Mach number decreases $(M=1.35$, and 5), leading to that strong perpendicular magnetics promote the regular to irregular transition, while moderate perpendicular magnetics suppress this transition compared to the corresponding hydrodynamic case. 
It is found that the transition from regular to irregular refraction is a complicated phenomenon and no clear trend, a more thorough investigation of the parameter space is left for the near future. 

Due to the large parameter space in MHD shock refraction, it is somewhat challenging to provide a complete 
taxonomy of all the irregular refraction patters. We present a sampling of the parameter space in which interesting irregular patterns occur. The self-similar transformation yields ``steady" solutions and hence enables us to focus our attention, in that we can concentrate the adaptive meshes where the refraction is taking place.  The numerical results show that there are two types for irregular MHD shock refraction, the first one is an MHD variant of single Mach reflexion $MRR$ with the appearance of a Mach stem.  
The second type is an MHD variant of convex-forwards irregular refraction $CFR$, with the reflected shock has apparently dispersed into a band of wavelets. 
The Mach stem and the band of wavelet emerging from Mach stem are classified as fast mode MHD shocks in MHD irregular refraction.
The vorticity deposited on the vicinity SC is transported on the the MHD waves for regular and irregular shock refractions, indicating that the mechanism by which a magnetic field suppresses the MHD RMI is valid independent of whether regular or irregular refraction occurs at the density interface, and whether a parallel or perpendicular magnetic field is present.
Moreover, the presence of magnetic field allows to weaken or completely vanish the second shear layer emerging from the triple-point between the incident shock, reflected fast shock and the Mach stem, indicating that the vorticity is transported from the vicinity of the density interface, preventing its instability. 
Perpendicular magnetic fields suppress more effectively this vorticity than the corresponding parallel magnetic fields.

\section*{Acknowledgement}
This research was supported by funding from King Abdullah University of Science and Technology (KAUST) under Grant No. BAS/1/1349-01-01.


\newpage
\section*{Reference}
\bibliography{apssamp}

\end{document}